\documentclass[aps,pra,twocolumn,superscriptaddress,nofootinbib]{revtex4-2}

\usepackage{amsmath, empheq,braket,amsthm,amssymb,graphics,amsfonts,array,color,subfigure, mathrsfs, dsfont, physics, float, bm}

\definecolor{myurlcolor}{rgb}{0,0,0.7}
\definecolor{myurlcolor1}{rgb}{0,0.7,0.1}
\definecolor{myrefcolor}{rgb}{0,0,0.7}

\usepackage[unicode=true,pdfusetitle, bookmarks=true,bookmarksnumbered=false,bookmarksopen=false, breaklinks=false,pdfborder={0 0 0},backref=false,colorlinks=true, linkcolor=myrefcolor,citecolor=myurlcolor1,urlcolor=myurlcolor]{hyperref}

\newtheorem{theorem}{Theorem}
\newtheorem*{theorem*}{Theorem}
\newtheorem{definition}{Definition}

\renewcommand{\selectlanguage}[1]{}

\begin{document}

\title{Majorization theoretical approach to entanglement enhancement via local filtration}

\author{Zacharie Van Herstraeten}
\affiliation{Wyant College of Optical Sciences, The University of Arizona, Tucson, Arizona, 85721, USA}
\author{Nicolas J. Cerf}
\affiliation{Centre for Quantum Information and Communication, \'{E}cole polytechnique de Bruxelles,  CP 165, Universit\'{e} libre de Bruxelles, 1050 Brussels, Belgium}
\affiliation{Wyant College of Optical Sciences, The University of Arizona, Tucson, Arizona, 85721, USA}



\author{Saikat Guha}
\affiliation{Department of Electrical and Computer Engineering, University of Maryland, College Park, Maryland, 20742, USA}
\affiliation{Wyant College of Optical Sciences, The University of Arizona, Tucson, Arizona, 85721, USA}

\author{Christos N. Gagatsos}
\affiliation{Department of Electrical and Computer Engineering, The University of Arizona, Tucson, Arizona, 85721, USA}
\affiliation{Wyant College of Optical Sciences, The University of Arizona, Tucson, Arizona, 85721, USA}
\affiliation{Program in Applied Mathematics, The University of Arizona, Tucson, Arizona 85721, USA}

\begin{abstract}
From the perspective of majorization theory, we study how to enhance the entanglement of a two-mode squeezed vacuum (TMSV) state by using local filtration operations. We present several schemes achieving entanglement enhancement with photon addition and subtraction, and then consider filtration as a general probabilistic procedure consisting in acting with local (non-unitary) operators on each mode.
From this, we identify a sufficient set of two conditions for these filtration operators to successfully enhance the entanglement of a TMSV state, namely the operators must be Fock-orthogonal (\textit{i.e.}, preserving the orthogonality of Fock states) and Fock-amplifying (\textit{i.e.}, giving larger amplitudes to larger Fock states).
Our results notably prove that ideal photon addition, subtraction, and any concatenation thereof always enhance the entanglement of a TMSV state in the sense of majorization theory.
We further investigate the case of realistic photon addition (subtraction) and are able to upper bound the distance between a realistic photon-added (-subtracted) TMSV state and a nearby state that is provably more entangled than the TMSV, thus extending entanglement enhancement to practical  schemes via the use of a notion of approximate majorization. Finally, we consider the state resulting from $k$-photon addition (on each of the two modes) on a TMSV state. We prove analytically that the state corresponding to $k=1$ majorizes any state corresponding to $2\leq k \leq 8$ and we conjecture the validity of the statement for all $k\geq 9$.
\end{abstract}

\maketitle

\section{Introduction}
\label{sec:intro}

\vspace{-0.2cm}
Quantum theory reveals the existence of correlations between quantum systems that are of a stronger essence than any classical counterpart.
As such, entanglement is a highly valuable quantum resource and is without surprise at the heart of many applications of quantum science.
From entanglement-based quantum key distribution \cite{Ekert1991-by}, to quantum sensing \cite{Quantum-Enhanced_Transmittance_Sensing, PhysRevLett.129.010501}, and entanglement-assisted communications \cite{Bennett2002-vt, Guha2020-cz}, quantum entanglement is used with great success to overtake the limits of what is possible in the classical world.

Entanglement is qualitatively refined into different categories depending on the nature of the carrier states.
Continuous-variable (CV) entanglement is the property of entangled states defined in an infinite-dimensional Hilbert space \cite{CerfBook,Braunstein2005-zw}.
It is the setting in which entanglement was originally thought, as in the (non-physical) state of two particles with perfectly correlated momenta~\cite{Einstein1935-zm}.
A realistic quantum-optical version of that CV state is found in the two-mode squeezed vacuum (TMSV), produced through spontaneous parametric down conversion.
The TMSV  state has non-perfect correlations distributed according to a Gaussian distribution, making this state part of the family of Gaussian states.
In a CV regime, a distinction should be made between Gaussian entanglement and non-Gaussian entanglement \cite{barral2023metrological}.
Gaussian entanglement can always be undone with passive linear optics (\textit{i.e.}, beam splitters and phase shifters) \cite{Walschaers2017-yo} and 
it cannot be distilled by means of Gaussian local operations and classical communication \cite{Eisert2002-sq, Fiurasek2002-vw, Giedke2002-vp}. Further, it cannot be used as a resource for quantum computational advantage \cite{Chabaud2023-bp}, a consequence of the fact that Wigner-positive states  are classically simulatable (with Wigner-positive measurements) \cite{Mari2012-fs}. 
Note that it can, in principle, achieve quantum advantage with Wigner-negative measurements, as in Gaussian boson sampling~\cite{Hamilton2017-mx}, though this advantage then stems from the measurement rather than the state itself. Therefore,
non-Gaussian entanglement remains viewed today as a crucial resource for quantum applications.

In practice, non-Gaussian entanglement is more challenging to produce than Gaussian entanglement.
The latter can indeed be produced by exploiting spontaneous parametric down conversion.
An effective strategy to generate non-Gaussian entanglement is to perform non-Gaussian operations on an entangled Gaussian state.
In particular, recent theoretical works have studied schemes of entanglement enhancement with photon addition and subtraction \cite{navarrete2012enhancing, Zhang2022-lo, superposition_addition_subraction, Das2016-xp}.

Quantifying the entanglement of mixed quantum states is in general a difficult task.
For such states, indeed, there exist a variety of different measures, ranging from logarithmic negativity, von Neumann or Rényi entropy of the reduced state, to squashed entanglement \cite{Horodecki2009-mk}.
The situation becomes however much simpler when it comes to pure entangled states, as most of these measures reduce to the entropy of entanglement.
In the asymptotic regime, the possibility to transform many copies of a pure entangled state into many copies of another one depends solely on the value of their respective entropy of entanglement. 
The single-shot (non-asymptotic) case, however, cannot be resolved with a single measure and requires more complex conditions, reflecting the absence of  regularization.
These conditions are encapsulated by Nielsen's theorem \cite{Nielsen1999-ee} which, for pure states, provides a necessary and sufficient condition for the transformation of one state into another via local operation and classical communication (LOCC).
Note that LOCC are in general non-reversible as they may include measurements \cite{Chitambar2014-yl}.

\begin{theorem}[Nielsen \cite{Nielsen1999-ee}]
Let $\ket{\Psi},\ket{\Phi}$ be pure states of a bipartite Hilbert space $\mathcal{H}=\mathcal{H}_1\otimes\mathcal{H}_2$.
We define $\hat{\sigma}_{\Psi}=\mathrm{Tr_2}[\ket{\Psi}\bra{\Psi}]$ and $\hat{\sigma}_{\Phi}=\mathrm{Tr_2}[\ket{\Phi}\bra{\Phi}]$.
Then, the state $\ket{\Psi}$ is transformable into $\ket{\Phi}$ via LOCC if and only if $\hat{\sigma}_{\Phi}\succ\hat{\sigma}_{\Psi}$ (\textit{i.e.}, $\hat{\sigma}_{\Phi}$ majorizes $\hat{\sigma}_{\Psi}$).
\end{theorem}

Nielsen's theorem calls for introducing two important concepts: (i) the Schmidt decomposition of bipartite pure states, and (ii) the theory of majorization. 
First (i), any pure bipartite state $\ket{\Psi}$ admits a Schmidt decomposition $\ket{\Psi}=\sum_n\sqrt{p_n}\ket{u_n}\ket{v_n}$, where $\lbrace p_n\rbrace$ are the components of a probability vector $\mathbf{p}$ and $\lbrace\ket{u_n}\rbrace$, $\lbrace\ket{v_n}\rbrace$ are two orthonormal vector sets.
The probabilities $\lbrace p_n\rbrace$ are called the Schmidt coefficients of $\ket{\Psi}$ and yield the eigenvalues of the partial-traced state (either over $\mathcal{H}_1$ or $\mathcal{H}_2$).
Second (ii), majorization theory is a mathematical framework used to compare disorder among probability distributions \cite{Marshall1979-oa}.
Two (infinite-dimensional) probability vectors $\mathbf{p},\mathbf{q}\in\mathbb{R}^{\mathbb{N}}$ obey the majorization relation $\mathbf{p}\succ\mathbf{q}$ if and only if there exists a column-stochastic matrix $\mathbf{D}$ such that $\mathbf{q}=\mathbf{D}\mathbf{p}$ (a column-stochastic matrix has non-negative entries, columns summing up to 1 and rows summing up to less or equal to 1 \cite{Kaftal2010-yv}).
In a similar fashion, we say that two density operators $\hat{\rho},\hat{\sigma}$ obey the majorization relation $\hat{\rho}\succ\hat{\sigma}$ if and only if their eigenvalues obey the corresponding majorization relation $\bm{\lambda}(\hat{\rho})\succ\bm{\lambda}(\hat{\sigma})$, where $\bm{\lambda}(\hat{\rho})$ is the vector of eigenvalues of $\hat{\rho}$. Note that if
both states have the same vector of eigenvalues (regardless of the ordering), we say that they are \textit{equivalent} and write it $\hat{\rho}\equiv\hat{\sigma}$.
This is a weaker condition than a strict equality between the two states, namely $\hat{\rho}=\hat{\sigma}$.
Let us mention that Nielsen's theorem was originally formulated in a finite-dimensional setting, and was later extended to CV entanglement \cite{Owari2008-ry}.


\begin{figure}[t]
\includegraphics[width=0.95\linewidth]{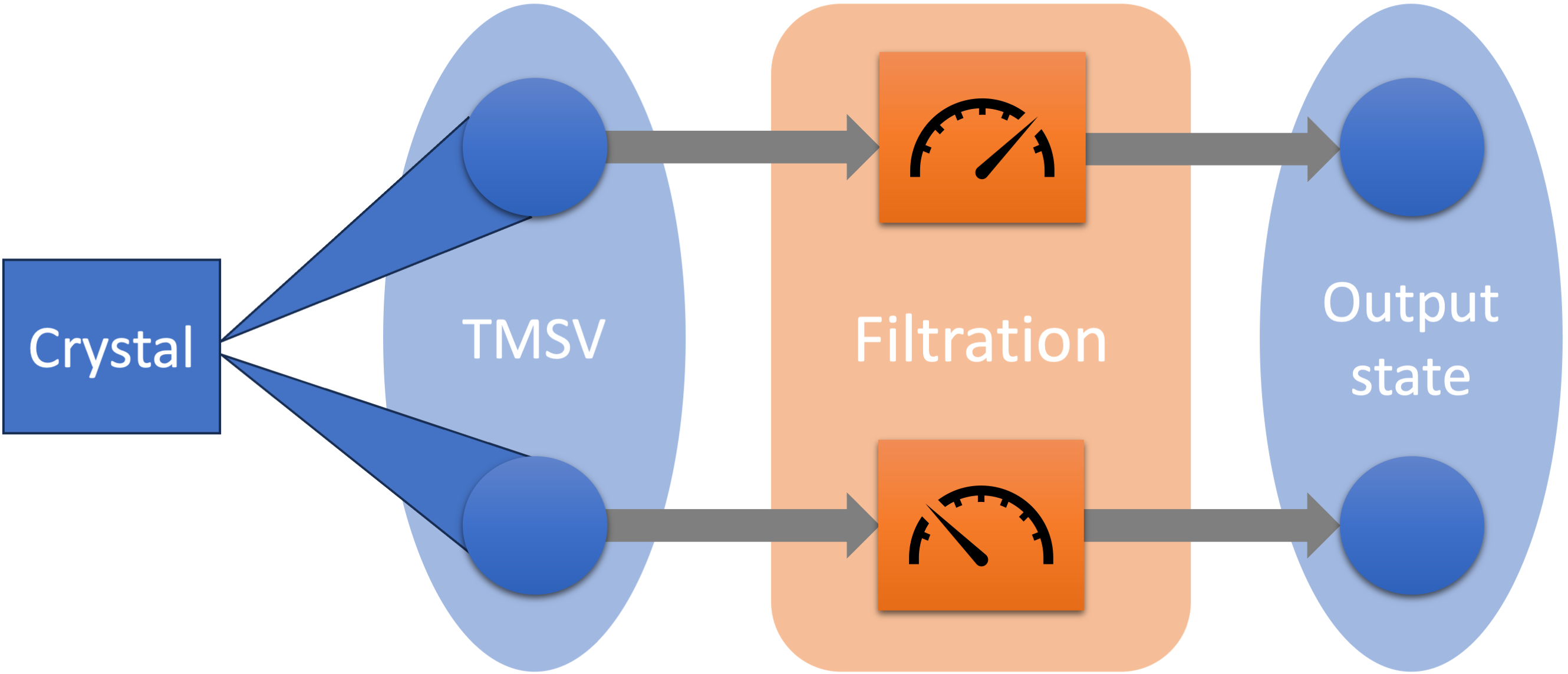}
\caption{
A pumped crystal produces a TMSV via spontaneous parametric down conversion.
Then, we act locally on both modes with some operators possibly including measurements.
We call that process \textit{filtration}.
We will study the conditions for filtration to produce an output state that is more entangled than the TMSV, in the majorization sense.
}
\label{fig:intro_setup}
\end{figure}

Throughout this paper, we will be interested in schemes of entanglement enhancement on a TMSV.
We consider a general setup where we are allowed to act locally on each mode of the TMSV with some local operations, possibly including measurements.
As such, these (non-unitary) operations are in general stochastic and associated to some probability of success.
We call that process of entanglement enhancement \textit{filtration}, as it keeps the resulting state conditionally on some measurement.
The setup is described on Fig. \ref{fig:intro_setup}.
We will then seek for conditions for the pair of operators to produce a state that is more entangled than the TMSV, in the strong sense of majorization (as per Nielsen's theorem).

At first glance, it may seem surprising that entanglement enhancement is possible by acting only locally on a quantum state.
This is explained by the fact that filtration operations are in general non-deterministic and are associated with a probability of success.
As such, filtration operations extend beyond the set of LOCCs \cite{Chitambar2014-yl}.
Once the distillable entanglement of the filtered state is multiplied by the probability of success, it cannot exceed the distillable entanglement of the original state.

This paper is structured as follows.
We use Sec.~\ref{sec:schemes} to define our notations and introduce several schemes of entanglement enhancement, going from basic to more general setups.
We end up with a similar scheme as the one described in Fig.~\ref{fig:intro_setup} and focus on this one in the following of the paper.
In Sec.~\ref{sec:results} we present our main analytical results.
We provide a sufficient condition for a pair of operator acting on a TMSV to produce a more entangled state.
That result allows us to prove that any concatenation of (ideal) creation or annihilation operators always enhance the entanglement of a TMSV.
Sec.~\ref{sec:realistic} is devoted to the cases of realistic photon addition and subtraction (with gain $g>1$ or transmittance $\eta<1$), which fall beyond the scope of the theorems proven in the previous section.
For these realistic operators, we provide an upper bound on the distance from the produced state to a state for which a majorization relation provably holds.
We finish by discussing our results and conclude in Sec.~\ref{sec:concl}.

\section{Entanglement enhancement schemes}
\label{sec:schemes}

In this section, we describe several schemes of entanglement enhancement on a TMSV.
Starting from a TMSV, we want to build a state that is more entangled.
Each scheme is defined by how we are allowed to interact with the TMSV.
We are going to describe three schemes for entanglement enhancement, with increasing complexity.
Then, in the next section, we will focus on the last of them, which is the more general.

Let us first define our notations.
We consider a bipartite Hilbert space $\mathcal{H}$ which is the tensor product of two infinite dimensional Hilbert spaces $\mathcal{H}_1$ (mode $1$) and $\mathcal{H}_2$ (mode $2$).
The annihilation operators on mode $1$ and mode $2$ are respectively $\hat{a}$ and $\hat{b}$; the creation operators on mode $1$ and $2$ are respectively $\hat{a}^\dagger$ and $\hat{b}^\dagger$.
They satisfy the canonical commutation relations $[\hat{a},\hat{a}^\dagger]=1$, $[\hat{b},\hat{b}^\dagger]=1$, and operators defined over different modes always commute.
For sake of simple notations, we define $\hat{A}(k)\vcentcolon=\hat{a}^{\dagger k}$ if $k\geq 0$ and $\hat{A}(k)\vcentcolon=\hat{a}^{-k}$ if $k<0$, so that the operator $\hat{A}(k)$ creates/annihilates $k$ photons on mode 1 (depending on the sign of $k$). 
We similarly define $\hat{B}(k)$ over mode 2.
Notice that $\hat{A}(k)$ and $\hat{B}(l)$ always commute, but $\hat{A}(k)$ and $\hat{A}(l)$ only commute when $k$ and $l$ have the same sign (similarly for $\hat{B}(k)$ and $\hat{B}(l)$).
We define the usual photon-number operators $\hat{n}_1\vcentcolon=\hat{a}^\dagger\hat{a}$ and $\hat{n}_2\vcentcolon=\hat{b}^\dagger\hat{b}$.
We will work in the Fock basis $\lbrace\ket{n}\rbrace$, where $\ket{n}$ is an eigenstate of the photon-number operator with eigenvalue $n\in\mathbb{N}$ (separately for $\mathcal{H}_1$ and $\mathcal{H}_2$, and then use the tensor product of these two bases for $\mathcal{H}$).
We use the vector norm $\Vert\ket{\psi}\Vert\vcentcolon=\sqrt{\bra{\psi}\ket{\psi}}$.

In each case, we consider a two-mode squeezer (TMS) fed by a two-mode vacuum input, so that the output is a TMSV.
The unitary operator of the TMS is defined as $\hat{U}_\lambda=\mathrm{exp}(r(\hat{a}^\dagger\hat{b}^\dagger-\hat{a}\hat{b}))$ with $r\in[0,\infty)$ and $\lambda=\tanh^2 r$.
Throughout this paper, we denote the TMSV state vector as $\ket{\Psi_\lambda}=\hat{U}_\lambda\ket{0,0}$, which gives: 
\begin{align}
    \ket{\Psi_\lambda}
    =
    \sqrt{1-\lambda}\,
    \sum\limits_{n=0}^{\infty}
    \sqrt{\lambda}^{\,n}\ket{n,n}.
\end{align}
The squeezing of the TMSV is measured in dB as $10\log_{10}e^{2r}=(20/\ln10)r=(20/\ln10)\tanh^{-1}(\sqrt{\lambda})$.
Partial tracing a TMSV over one of its modes yields a thermal state $\hat{\tau}_\lambda=\mathrm{Tr}_2[\ket{\Psi_\lambda}\bra{\Psi_\lambda}]$, which gives:
\begin{align}
    \hat{\tau}_{\lambda}
    =
    (1-\lambda)
    \sum\limits_{n=0}^{\infty}
    \lambda^n
    \ket{n}\bra{n},
\end{align}
and has mean photon-number $\lambda/(1-\lambda)=\sinh^2 r$.
The vector of eigenvalues of the thermal state $\hat{\tau}_\lambda$ is $\bm{\tau}$ with components $\tau_n=(1-\lambda)\lambda^n$.
The vector $\bm{\tau}$ is also the vector of Schmidt coefficients of the TMSV $\ket{\Psi_\lambda}$.

The different schemes that we are going to define hereafter enable us to act on $\ket{\Psi_\lambda}$ in order to create  a pure state $\ket{\Phi}$ (heralded by the success of the filtration) that is \textit{more} entangled than $\ket{\Psi_\lambda}$ in the sense of majorization theory.  
We say that a filtration protocol enhances the entanglement when the final state can be transformed back into the original state via LOCC.
We know from Nielsen theorem \cite{Nielsen1999-ee} that $\ket{\Phi}$ is transformable into $\ket{\Psi_\lambda}$ via LOCC if and only if $\hat{\tau}_\lambda\succ\hat{\sigma}_{\Phi}$, where $\hat{\sigma}_{\Phi}=\mathrm{Tr}_2[\ket{\Phi}\bra{\Phi}]$ is the reduced single-mode state associated with $\ket{\Phi}$. Thus, checking whether the thermal state $\hat{\tau}_\lambda$ majorizes the reduced single-mode state after filtration is how we will assess that filtration has been successful.


The present section is solely focused on defining the filtration schemes under investigation. We will study how entanglement is enhanced in the next section.
We make a distinction between schemes and setups as follows: a scheme is a family of setups (with multiple possible parameters), whereas a setup is a particular instance that produces a well defined state. 
We consider that two setups are equivalent when they produce two states having same Schmidt coefficients (hence, two states that are equivalent in the sense of majorization theory), since in that case it is possible to transform each state into the other one using LOCC.

\subsection*{(a) Dual-mode single addition or subtraction}

\begin{figure}[t]
\includegraphics[width=1\linewidth]{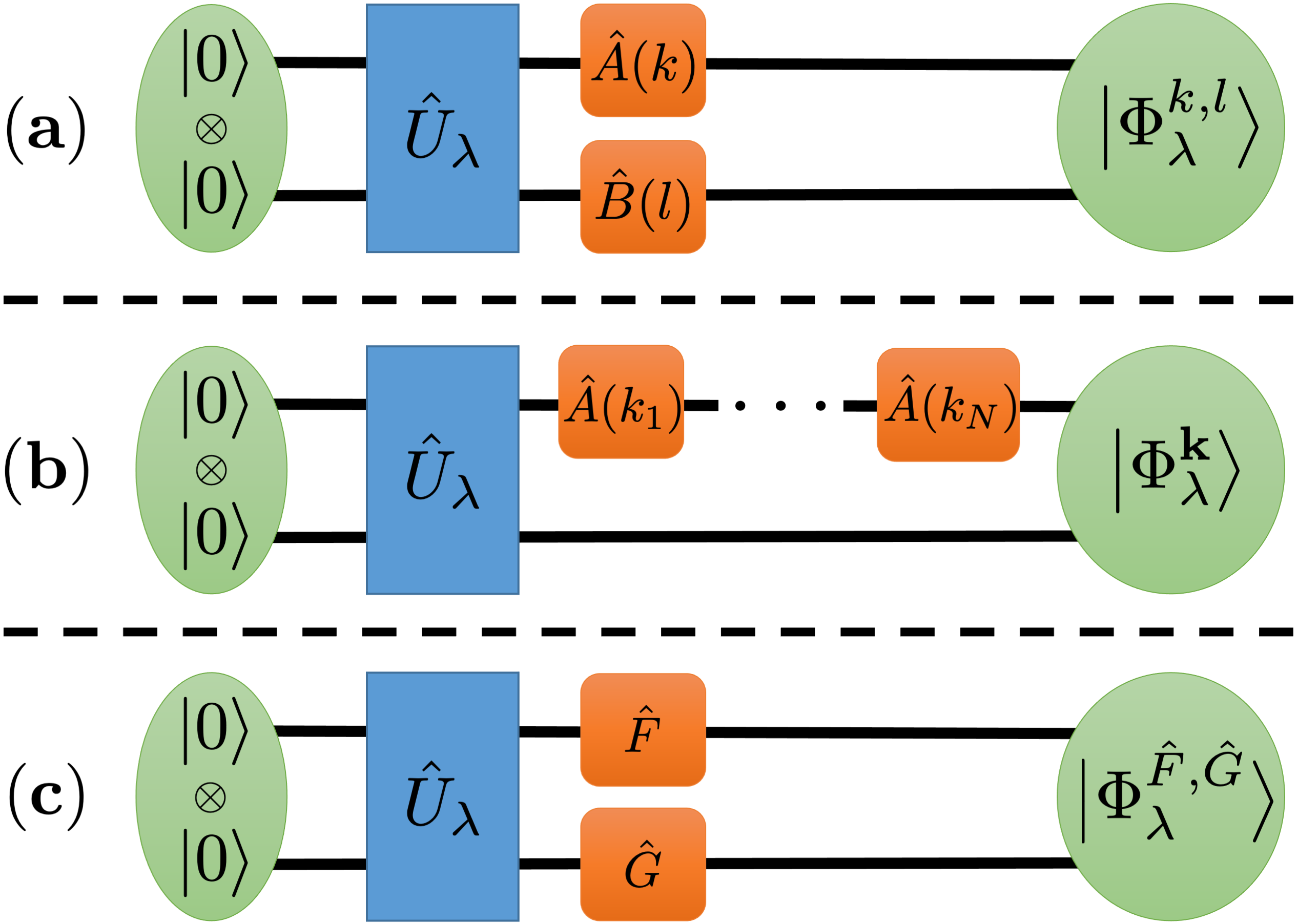}
\caption{Illustration of the 3 different schemes considered in Section \ref{sec:schemes}.
Each (orange) round-cornered box is an ideal photon addition/subtraction ($k,l\in\mathbb{Z}$), or a more general single-mode quantum operator ($\hat{F},\hat{G}$).
Scheme (a) corresponds to dual-mode single addition/subtraction (see also Ref. \cite{navarrete2012enhancing}) and yields the pure state $\big\vert\Phi_\lambda^{k,l}\big\rangle$.
Scheme (b) corresponds to single-mode multi addition/subtraction and yields the pure state $\big\vert\Phi_\lambda^{\mathbf{k}}\big\rangle$.
Finally, scheme (c) corresponds to the filtration scheme and yields the pure state $\smash{\big\vert\Phi^{\hat{F},\hat{G}}_\lambda\big\rangle}$.
Note that, for each scheme, we consider that the output state $\ket{\Phi}$ is normalized.
Scheme (c) is a generalization of scheme (b), which itsef generalizes scheme (a).
}
\label{fig:schemes}
\end{figure}

The first scheme that we consider allows us to add or subtract an arbitrary number of photons once for each mode of the TMSV.
This means that a setup from that scheme is uniquely defined by two integers $k,l\in\mathbb{Z}$, where $k$ (resp. $l$) is the number of added/subtracted photons on mode 1 (resp. 2).
This scheme is similar to the one proposed in Ref. \cite{navarrete2012enhancing}.
See Fig. \ref{fig:schemes}(a) for an illustration.
The pure state resulting from acting on the TMSV with a setup of this scheme is then:
\begin{align}
    \big\vert\Phi_\lambda^{k,l}\big\rangle
    \vcentcolon=
    \mathcal{N}_{k,l}^{\, -1/2}
    \hat{A}(k)\hat{B}(l)
    \ket{\Psi_\lambda},    
    \label{eq:phi_dualmode_single}
\end{align}
with   $\mathcal{N}_{k,l}\vcentcolon=\Vert\hat{A}(k)\hat{B}(l)\ket{\Psi_\lambda}\Vert^2$ being a normalization constant.
Note that $\vert\Phi^{0,0}_\lambda\rangle=\vert\Psi_\lambda\rangle$.
The corresponding single-mode state is then $\hat{\sigma}^{k,l}_{\lambda}\vcentcolon=\mathrm{Tr}_2[\vert\Phi^{k,l}_\lambda\rangle\langle\Phi^{k,l}_\lambda\vert]$, which should be compared to $\hat{\tau}_\lambda$.

Hereafter, we identify some 
equivalences in this entanglement enhancement scheme:


\begin{itemize}
    \item \textit{Mode exchange equivalence}\\
    From the invariance of the TMSV state under the exchange of the two modes, it follows that swapping operations performed on mode 1 and mode 2 yields another state which is simply the original state with mode 1 and 2 swapped.
    Then, we know from Schmidt decomposition that both single-mode reduced states are equivalent.
    We write:
    \begin{align}
    \hat{\sigma}_{\lambda}^{k,l}\equiv\hat{\sigma}_{\lambda}^{l,k}
    \qquad\forall k,l\in\mathbb{Z}
    \label{eq:mode_swapping_eq}
    \end{align}
    
    \item \textit{Dual addition vs. subtraction equivalence}\\
    Ref. \cite{navarrete2012enhancing} shows that adding or subtracting $k$ photons on both modes of the TMSV yields two equivalent states (they are different but have the same eigenvalues).
    We write:
    \begin{align}
    \hat{\sigma}^{k,k}_\lambda\equiv\hat{\sigma}^{-k,-k}_\lambda
    \qquad\forall k\in\mathbb{Z}
    \label{eq:pairwise_addsub_eq}
    \end{align}
    \item \textit{Single addition vs. subtraction identity}\\
    Ref. \cite{navarrete2012enhancing} proves the remarkable relation $ \hat{b}\ket{\Psi_\lambda}=\sqrt{\lambda}\, \hat{a}^{\dagger}\ket{\Psi_\lambda}$,
      which allows to move a creation operator on one mode into an annihilation operator on the other mode. We write:
\begin{align}
\hat{\sigma}_{\lambda}^{0,k}=\hat{\sigma}_{\lambda}^{-k,0}
    \qquad\forall k\in\mathbb{Z}
\end{align}
where we actually have a strict equality instead of an equivalence (the extra $\lambda$ factor disappears after normalization). The superscripts zero are crucial here because the above relation only holds if acting on the TMSV state $\ket{\Psi_\lambda}$. The use of this relation is further discussed in Appendix \ref{apd:TMS-commutation-identity}.
\end{itemize}

\subsection*{(b) Single-mode multiple addition or subtraction}

In light of this observation (see also Appendix \ref{apd:TMS-commutation-identity}), we are now going to consider a scheme where we apply an arbitrary number of creation/annihilation operators, but on only one of the two modes.
A setup of this new scheme is then defined by a vector $\mathbf{k}\in\mathbb{Z}^N$, where $N$ is the total number of creation/annihilation operators (each associated to a $k_n$).
See Fig. \ref{fig:schemes}(b) for an illustration.
This corresponds to applying the following operator $\hat{F}_{\mathbf{k}}$ on mode $1$ of the TMSV:
\begin{align}
    \hat{F}_{\mathbf{k}}
    =
    \hat{A}(k_N)
    \hat{A}(k_{N-1})
    \cdots
    \hat{A}(k_2)
    \hat{A}(k_1).
    \label{eq:concatenation_addsub}
\end{align}
Note that since $\hat{A}(k)$ and $\hat{A}(l)$ do not commute in general, the order introduced in Eq. \eqref{eq:concatenation_addsub} is important (first photon addition/subtraction is $k_1$, second is $k_2$, and so on).
We may then define the resulting pure state of operator $\hat{F}_{\mathbf{k}}$ acting on $\ket{\Psi_\lambda}$ as follows:
\begin{align}
    \vert\Phi_\lambda^{\mathbf{k}}\rangle
    \vcentcolon=
    \mathcal{N}_{\mathbf{k}}^{\, -1/2}\hat{F}_\mathbf{k}\ket{\Psi_\lambda},
\end{align}
where $\hat{F}_{\mathbf{k}}$ acts on mode 1 and the normalization constant is defined as $\mathcal{N}_{\mathbf{k}}\vcentcolon=\Vert\hat{F}_{\mathbf{k}}\ket{\Psi_\lambda}\Vert^2$.

From Eq. \eqref{eq:add-sub-id} (or its corollary), observe that $\vert\Phi_\lambda^{k,l}\rangle=\vert\Phi^{\mathbf{k}}_\lambda\rangle$ if we choose $\mathbf{k}=(-l,k)$.
This demonstrates that the single-mode multiple addition/subtraction scheme is more general than the dual-mode single addition/subtraction scheme.

\subsection*{(c) General filtration scheme}

We finally introduce a last scheme, the most general one that we will consider.
It generalizes the two formerly introduced schemes.
See Fig. \ref{fig:schemes}(c) for an illustration.
We act on the first mode of the TMSV with a filtration operator $\hat{F}$ and on its second mode with another filtration operator $\hat{G}$, producing hence the following pure state:
\begin{align}
    \big\vert\Phi^{\hat{F},\hat{G}}_\lambda\big\rangle
    &=
    \mathcal{N}^{\, -1/2}_{\hat{F},\hat{G}}
    \big(\hat{F}\otimes\hat{G}\big)\ket{\Psi_\lambda}
\end{align}
where the normalization constant is defined as $\mathcal{N}_{\hat{F},\hat{G}}\vcentcolon=\Vert(\hat{F}\otimes\hat{G})\ket{\Psi_\lambda}\Vert^2$.

The scheme considered here is very general, but it is important to highlight an implicit underlying assumption.
By considering that we act on both mode with some operators ($\hat{F}$, $\hat{G}$), we imply that we transform pure states into (possibly non-normalized) pure states.
For that reason, quantum channels applying pure states onto mixed states do not belong to the present scheme. Note, also, that for the filtration to succeed with a non-zero probability, the filtration operators $\hat{F}$, $\hat{G}$ must be bounded. This is not the case for the $\hat{A}(k)$ operators (with $k\ne 0$), for example, which are unbounded. This is the reason why we will be referring to \textit{ideal} photon addition/subtraction (whose success probability is strictly zero), in contrast to \textit{realistic} photon addition/subtraction (see Sec.
\ref{sec:realistic}).

\section{Main theorems}
\label{sec:results}

In this section, we present our main analytical results.
We consider the last scheme introduced in Sec. \ref{sec:schemes}(c), see Fig. \ref{fig:schemes}(c).
First, we look at the simpler case of only one filtration operator acting on mode 1 (see Theorem \ref{theorem:single_mode}), then we will move to the general case of two filtration operators acting separately on mode 1 and mode 2 (see Theorem \ref{theorem:dual_mode}).
For both cases, we will identify a set of sufficient conditions for these operators to produce a state that is more entangled than the original TMSV state, using Nielsen's theorem (see Introduction, Sec. \ref{sec:intro}) to compare entangled pure states.
Our proof relies on the explicit construction of a column-stochastic matrix relating the Schmidt coefficients of both bipartite entangled pure states.

As an introduction to the present section, let us define two properties of a filtration operator that will play an important role in the proof of our theorems.

\begin{definition}[Fock-orthogonal operator]
An operator $\hat{F}$ is Fock-orthogonal iff it preserves the orthogonality of Fock states, \textit{i.e.} $\langle\hat{F}m\vert\hat{F}n\rangle=0, \ \forall m\neq n$.
\label{def:fock_ortho}
\end{definition}
We use the compact notation $\vert\hat{F}n\rangle\vcentcolon=\hat{F}\ket{n}$.
This property can be understood as a relaxed form of unitarity: it requires orthogonality conservation over the basis of Fock states; but it is not necessarily true for arbitrary bases. 
For that reason, Fock-orthogonal operators are in general non-unitary.
In fact, non-unitarity is a crucial property for our local scheme to enhance entanglement because any local unitary operator has no influence on entanglement. 
The next property is precisely related to non-unitarity.

\begin{definition}[Fock-amplifying operator]
An operator $\hat{F}$ is Fock-amplifying iff it gives greater amplitudes to higher Fock states, \textit{i.e.} $\Vert\hat{F}\ket{n}\Vert\leq\Vert\hat{F}\ket{n+1}\Vert, \ \forall n.$
\label{def:fock_amp_op}
\end{definition}
In some sense, the Fock-amplifying condition hints to the fact that the filtration operator is generally not unitary.
Indeed, for any unitary operator $\hat{U}$, we have $\Vert\hat{U}\ket{n}\Vert=1$ so that the inequality $\Vert\hat{U}\ket{n}\Vert\leq\Vert\hat{U}\ket{n+1}\Vert$ is trivially satisfied.
Thus, as soon as one of the inequalities is strict, the operator must necessarily be non-unitary.

Interestingly, the conditions for an operator $\hat{F}$ to be Fock-orthogonal or Fock-amplifying boil down to conditions on the operator $\hat{F}^\dagger\hat{F}$.
Indeed, the operator $\hat{F}$ is Fock-orthogonal if and only if the operator $\hat{F}^\dagger\hat{F}$ is diagonal in the Fock basis, and the operator $\hat{F}$ is Fock-amplifying if and only if the diagonal entries of $\hat{F}^\dagger\hat{F}$ in the Fock basis are non-decreasing with $n$.
Finally, the operator $\hat{F}$ is both Fock-orthogonal and Fock-amplifying if and only if $\hat{F}^\dagger\hat{F}=f(\hat{n})$ where $f:\mathbb{N}\to\mathbb{R}_+$ is a non-decreasing function.

From Defs. \ref{def:fock_ortho} and \ref{def:fock_amp_op}, it is clear that an operator $\hat{F}$ is both Fock-orthogonal and Fock-amplifying if and only if it acts onto the Fock basis as follows:
\begin{align}
    \hat{F}\ket{n}
    =
    \varphi_n\ket{\phi_n},
    \label{eq:fock_amp_canonical}
\end{align}
where $\lbrace\ket{\phi_n}\rbrace$ is an orthonormal set and $\bm{\varphi}\in\mathbb{C}^{\mathbb{N}}$ is an amplitude vector such that $\abs{\varphi_n}\leq\abs{\varphi_{n+1}}\ \forall n$.
The operator $\hat{F}$ is uniquely defined by $\lbrace\ket{\phi_n}\rbrace$ and $\bm{\varphi}$.
Table~\ref{table:examples_fock_amp} mentions a few examples of common Fock-orthogonal Fock-amplifying filtration operators.

\renewcommand{\arraystretch}{1.5}
\begin{table}[t]
	\resizebox{1\columnwidth}{!}{%
	\begin{tabular}{|l|c|c|c|}
        \hline
        Operator      & $\quad\hat{F}\quad$             & $\varphi_n$              & $\lbrace\ket{\phi_n}\rbrace$ \\ \hline\hline
        Annihilation   & $\hat{a}^k$           & $\;\sqrt{n!/(n-k)!}\,[n\geq k]\;$ & $\;\lbrace \ket{n-k}\rbrace\;$   \\ \hline
        Creation      & $\hat{a}^{\dagger k}$ & $\sqrt{(n+k)!/n!}$ & $\lbrace \ket{n+k}\rbrace$   \\ \hline
        Photon-number & $\hat{n}$             & $n$                & $\lbrace \ket{n}\rbrace$     \\ \hline
        NLA           & $g^{\hat{n}}$         & $g^n$              & $\lbrace \ket{n}\rbrace$     \\ \hline
        \end{tabular}
	}
\caption{
Examples of Fock-orthogonal Fock-amplifying filtration operators, where the amplitudes $\varphi_n$ and vector sets $\lbrace\ket{\phi_n}\rbrace$ refer to Eq. \eqref{eq:fock_amp_canonical}. The (ideal) noiseless linear amplifier (NLA) is associated to a gain $g\geq 1$ \cite{Fiurasek2012-ev, Micuda2012-sw, Gagatsos2012-lo}. 
Note that in the case of the annihilation operator, some vectors $\ket{\phi_n}=\ket{n-k}$ are ill-defined; this is not a problem since they are associated to a zero amplitude (cf. the indicator function~$[.]$). 
Finally, observe that all of the above-mentioned operators are Fock-preserving (see Def. \ref{def:fock-preserving}), so that every concatenation thereof yields another Fock-preserving Fock-orthogonal Fock-amplifying operator (from Theorem \ref{th:concat}).}
\label{table:examples_fock_amp}
\end{table}

\subsection*{Single-mode majorization theorem}

With these newly defined properties, we are now in position to present our first theorem.
In this subsection, we consider an entanglement-enhancement scheme where we only act on a TMSV on its first mode with some operator $\hat{F}$.
This corresponds to the scheme of Fig. \ref{fig:schemes}\textrm{(c)} with $\hat{G}$ chosen to the identity operator $\hat{\mathds{1}}$.

\begin{theorem}
Let $\hat{F}$ be a Fock-orthogonal and Fock-amplifying operator.
Let $\hat{\tau}$ be a thermal state, and $\hat{\sigma}$ the result of $\hat{F}$ acting on $\hat{\tau}$, so that $\hat{\sigma}=\hat{F}\hat{\tau}\hat{F}^{\dagger}/\mathrm{Tr}[\hat{F}\hat{\tau}\hat{F}^{\dagger}]$.
Then, the majorization relation $\hat{\sigma}\prec\hat{\tau}$ holds.
\label{theorem:single_mode}
\end{theorem}

\begin{proof}
We define the vector $\bm{\tau}\in\mathbb{R}^{\mathbb{N}}$ as the vector of eigenvalues of the thermal state $\hat{\tau}$, so that $\tau_n=(1-\lambda)\lambda^n$.
Then, using Eq. \eqref{eq:fock_amp_canonical}, the density operator $\hat{\sigma}$ can be expressed as follows:
\begin{align}
    \hat{\sigma}
    =
    \mathcal{N}^{-1}
    \sum\limits_{n=0}^{\infty}
    \tau_n\abs{\varphi_{n}}^2
    \ket{\phi_n}\bra{\phi_n}
\end{align}
where the normalization constant is $\mathcal{N}=\mathrm{Tr}[\hat{F}\hat{\tau}\hat{F}^\dagger]=\sum_n\tau_n\abs{\varphi_n}^2$.
It follows from the Fock-orthogonality of $\hat{F}$ that the set $\lbrace\ket{\phi_n}\rbrace$ is orthonormal, so that the vector $\bm{\sigma}\in\mathbb{R}^\mathbb{N}$ with components $\sigma_n\vcentcolon=\mathcal{N}^{-1}\tau_n\abs{\varphi_n}^2$ is the vector of eigenvalues of $\hat{\sigma}$.
Then, it suffices to find a column-stochastic matrix $\mathbf{D}$ such that $\bm{\sigma}=\mathbf{D}\bm{\tau}$ in order to prove that $\hat{\sigma}\prec\hat{\tau}$.

To build such a column-stochastic matrix, we focus on a particular structure.
We consider a matrix $\mathbf{D}$ that is lower triangular and circulant (repeating the columns with an increasing offset), as follows:
\begin{align}
    \mathbf{D}=
    \begin{pmatrix}
        d_0 & 0 & 0 & 0 &\cdots\\
        d_1 & d_0 & 0 & 0 &\cdots\\
        d_2 & d_1 & d_0 & 0 &\cdots\\
        d_3 & d_2 & d_1 & d_0 &\cdots\\
        \vdots &\vdots &\vdots &\vdots &\ddots
    \end{pmatrix}.
    \label{eq:structure_columnstocastic}
\end{align}
The components of $\mathbf{D}$ are defined from the vector $\mathbf{d}$ as
$D_{ij}=d_{i-j}$, with the convention that $d_n=0$, $\forall n<0$.
A motivation for this choice of structure is that it is an easy task to build a column-stochastic matrix in this way.
Indeed, $\mathbf{D}$ is column-stochastic if and only if the vector $\mathbf{d}$ is a probability vector ($\sum_n d_n=1$ and $d_n\geq 0\ \forall n$).
With the additional constraint that $\bm{\sigma}=\mathbf{D}\bm{\tau}$, we find the components of $\mathbf{d}$, so that $d_n=\lambda^n(\abs{\varphi_n}^2-\abs{\varphi_{n-1}}^2)/\mathcal{N}$ with the convention that $\varphi_{n}=0,\ \forall n<0$.
Indeed, this gives:
\begin{align}
    \sum\limits_{j=0}^{\infty} D_{ij}\tau_j
    &=
    \sum\limits_{j=0}^{i} d_{i-j}(1-\lambda)\lambda^j
    \nonumber
    \\&=
    \frac{1}{\mathcal{N}}(1-\lambda)\sum\limits_{j=0}^{i} \lambda^{i}\left(\abs{\varphi_{i-j}}^2-\abs{\varphi_{i-j-1}}^2\right)
    \nonumber
    \\&=
    \frac{1}{\mathcal{N}}\tau_i\abs{\varphi_{i}}^2=\sigma_i.
    \nonumber
\end{align}
Let us now check that $\mathbf{d}$ is a probability vector.
It is normalized to $1$:
\begin{align*}
    \sum\limits_{n=0}^{\infty}
    d_n
    &=
    \frac{1}{\mathcal{N}}
    \sum\limits_{n=0}^{\infty}
    \lambda^n\abs{\varphi_n}^2
    -
    \frac{1}{\mathcal{N}}
    \sum\limits_{n=0}^{\infty}
    \lambda^n\abs{\varphi_{n-1}}^2
    \\
    &=
    \frac{1}{\mathcal{N}}
    \sum\limits_{n=0}^{\infty}
    \lambda^n\abs{\varphi_n}^2
    -
    \frac{1}{\mathcal{N}}
    \sum\limits_{n=0}^{\infty}
    \lambda^{n+1}\abs{\varphi_{n}}^2
    \\
    &=
    \frac{1}{\mathcal{N}}(1-\lambda)
    \sum\limits_{n=0}^{\infty}
    \lambda^n\abs{\varphi_n}^2
    =
    \frac{1}{\mathcal{N}}\sum\limits_{n=0}^{\infty}\tau_n\abs{\varphi_n}^2
    =1,
\end{align*}
where the last equality follows from the definition of the normalization constant $\mathcal{N}$.
Then, the components of $\mathbf{d}$ are non-negative as soon as $\abs{\varphi_n}-\abs{\varphi_{n-1}}\geq 0\ \forall n$, which follows from the assumption that $\hat{F}$ is Fock-amplifying.
The matrix $\mathbf{D}$ is thus column-stochastic and such that $\bm{\sigma}=\mathbf{D}\bm{\tau}$.
This proves $\hat{\sigma}\prec\hat{\tau}$.
\end{proof}

\subsection*{Dual-mode majorization theorem}

Theorem \ref{theorem:single_mode} considers a scheme where we only interact with one mode of the TMSV.
Our next theorem is a generalization to the case where we use two operators to act on each mode of the TMSV.
The setup is illustrated in Fig. \ref{fig:schemes}(c).
An important ingredient of this second theorem is a property closely related to Fock-amplifying operators, but for a pair of operators.

\begin{definition}[Jointly Fock-amplifying operator pair]
A pair of operators $(\hat{F},\hat{G})$ is jointly Fock-amplifying iff
\begin{align}
    \Vert\hat{F}\ket{n}\Vert\cdot
    \Vert\hat{G}\ket{n}\Vert
    \leq
    \Vert\hat{F}\ket{n+1}\Vert\cdot
    \Vert\hat{G}\ket{n+1}\Vert
    \quad\forall n.
    \nonumber
\end{align}
\end{definition}
A jointly Fock-amplifying pair of operators gives a greater amplitude to higher Fock states.
From the definition, it is obvious that any two Fock-amplifying operators form a jointly Fock-amplifying pair.
However, two operators do not need to be separately Fock-amplifying in order to make a jointly Fock-amplifying pair.
As an example, a noiseless linear amplifier with gain $g$ and a noiseless linear attenuator \cite{Fiurasek2012-ev, Micuda2012-sw, Gagatsos2012-lo} with transmittance $\eta$ make a Fock-amplifying pair as soon as $g\cdot\eta\geq 1$, even though the noiseless linear attenuator is not Fock-amplifying ($\forall \eta$).
Of course, any pair $(\hat{F},\hat{\mathds{1}})$ is jointly Fock-amplifying if and only if $\hat{F}$ is Fock-amplifying ($\hat{\mathds{1}}$ is the identity operator).

\begin{theorem}
    Let $\hat{F},\hat{G}$ be two Fock-orthogonal operators such that $(\hat{F},\hat{G})$ is jointly Fock-amplifying. Let $\ket{\Psi}$ be a TMSV state, with the associated thermal state $\hat{\tau}=\mathrm{Tr}_2[\ket{\Psi}\bra{\Psi}]$, and let $\hat{\sigma}$ be the single-mode result of $(\hat{F},\hat{G})$ acting on $\ket{\Psi}$, such as
    \begin{align}
    \hat{\sigma}=
    \mathcal{N}^{-1}
    \Tr_2[(\hat{F}\otimes\hat{G})
    \ket{\Psi}\bra{\Psi}
    (\hat{F}^\dagger\otimes\hat{G}^\dagger)]
    \end{align}
    where $\mathcal{N}=\Vert(\hat{F}\otimes\hat{G})\ket{\Psi}\Vert^2$.
    Then, the majorization relation $\hat{\sigma}\prec\hat{\tau}$ holds.
    \label{theorem:dual_mode}
\end{theorem}

\begin{proof}
Let the vector $\bm{\tau}$ with components $\tau_n$ be the vector of eigenvalues of $\hat{\tau}$.
Since $\hat{F}$ and $\hat{G}$ are Fock-orthogonal, we can write their action onto Fock states as respectively $\hat{F}\ket{n}=\varphi_n\ket{\phi_n}$ and $\hat{G}\ket{n}=\gamma_n\ket{\psi_n}$, where $\lbrace\ket{\phi_n}\rbrace$ and $\lbrace\ket{\psi_n}\rbrace$ are two orthonormal sets.
The two-mode pure state $\ket{\Phi}=\mathcal{N}^{-1/2}(\hat{F}\otimes\hat{G})\ket{\Psi}$ is then:
\begin{align}
    \ket{\Phi}
    =
    \mathcal{N}^{-1/2}\,
    \sum\limits_{n=0}^{\infty}
    \sqrt{\tau_n}\,
    \varphi_n\gamma_n
    \ket{\phi_n}\ket{\psi_n}.
\end{align}
The above expression is the Schmidt decomposition of $\ket{\Phi}$, from the orthogonality of the sets $\lbrace\ket{\phi_n}\rbrace$ and $\lbrace\ket{\psi_n}\rbrace$.
The single-mode state $\hat{\sigma}$ is then computed as:
\begin{align}
    \hat{\sigma}
    =
    \mathcal{N}^{-1}\,
    \sum\limits_{n=0}^{\infty}\,
    \tau_n\,
    \abs{\varphi_n}^2\,
    \abs{\gamma_n}^2\,
    \ket{\phi_n}\bra{\phi_n}.
\end{align}
Let us now define the operator $\hat{D}$ acting on the Fock basis as $\hat{D}\ket{n}=\varphi_n\gamma_n\ket{\phi_n}$, and observe that $\hat{\sigma}=\hat{D}\hat{\tau}\hat{D}^\dagger/\mathrm{Tr}[\hat{D}\hat{\tau}\hat{D}^\dagger]$.
Since the set $\lbrace\ket{\phi_n}\rbrace$ is orthonormal, $\hat{D}$ is Fock-orthogonal.
Since $(\hat{F},\hat{G})$ is a jointly Fock-amplifying pair, it follows that $\abs{\varphi_n}\abs{\gamma_n}\leq\abs{\varphi_{n+1}}\abs{\gamma_{n+1}}$, so that $\hat{D}$ is Fock-amplifying.
From Theorem \ref{theorem:single_mode}, this then implies that $\hat{\sigma}\prec\hat{\tau}$.
\end{proof}

\subsection*{Concatenation theorem}

The last question we address in this section concerns the concatenation of filtration operators: is the concatenation of two Fock-orthogonal (resp. Fock-amplifying) operators also Fock-orthogonal (resp. Fock-amplifying)?
Observe first that if we have two operators $\hat{F}$ and $\hat{G}$ which are both Fock-orthogonal and Fock-amplifying, their concatenation $\hat{G}\hat{F}$ may itself be neither.
As an example, consider $\hat{F}=\hat{U}\hat{a}$ and $\hat{G}=\hat{a}$, where $\hat{U}$ is some unitary operator. It is easily seen that both $\hat{F}$ and $\hat{G}$ are Fock-orthogonal and Fock-amplifying; however, their concatenation $\hat{G}\hat{F}=\hat{a}\hat{U}\hat{a}$ is in general not.
In this subsection, we are going to identify an additional property which ensures that the Fock-orthogonal or Fock-amplifying property is preserved under concatenation.

\begin{definition}[Fock-preserving operator]
An operator $\hat{F}$ is Fock-preserving iff: {\bf (a)}~it maps Fock states onto Fock states, \textit{i.e.}, $\langle k\vert\hat{F}n\rangle\langle l\vert\hat{F}n\rangle=0,\ \forall n,\ k\neq l$, and {\bf(b)}~it maps higher Fock states onto higher Fock states, \textit{i.e.}, $\langle k\vert\hat{F}m\rangle\langle l\vert\hat{F}n\rangle=0,\forall k>l,\ m< n$.
\label{def:fock-preserving}
\end{definition}
Let us give more intuition about Def. \ref{def:fock-preserving} by introducing the matrix $\mathbf{F}$ with elements $F_{ij}=\bra{i}\hat{F}\ket{j}$.
Condition \textrm{(a)} of Def. \ref{def:fock-preserving} implies that every column of $\mathbf{F}$ has at most one non-zero entry.
Condition \textrm{(b)} of Def. \ref{def:fock-preserving} implies that the row-index of non-zero entries must be non-decreasing as the column-index increases.
From these observations, we understand that an operator $\hat{F}$ is Fock-preserving if and only if it acts on the Fock basis as:
\begin{align}
    \hat{F}\ket{n}
    =
    \varphi_n\ket{m_n}
    \label{eq:fock-concatenable}
\end{align}
where $\bm{\varphi}\in\mathbb{C}^\mathbb{N}$ is an amplitude vector and $\mathbf{m}\in\mathbb{N}^\mathbb{N}$ is a vector with integer non-decreasing components ($m_n\leq m_{n+1}$).
In Eq. \eqref{eq:fock-concatenable}, the Fock state $\ket{n}$ is applied onto another Fock state with photon-number $m_n$.

\begin{theorem}[Concatenability]
    \label{th:concat}
    {\bf{(a)}}
    If $\hat{F},\hat{G}$ are Fock-preserving, then $\hat{G}\hat{F}$ is Fock-preserving.
    {\bf(b)}
    If $\hat{F}$, $\hat{G}$ are Fock-orthogonal, then $\hat{G}\hat{F}$ is Fock-orthogonal provided that $\hat{F}$ is Fock-preserving too.
    {\bf{(c)}}
    If $\hat{F}$, $\hat{G}$ are Fock-amplifying, then $\hat{G}\hat{F}$ is Fock-amplifying provided that $\hat{F}$ is Fock-preserving too.
\end{theorem}

The proof of Theorem \ref{th:concat} is provided in Appendix \ref{apd:proofs}.
In a nutshell, the Fock-preserving property allows an operator to carry over its Fock-orthogonal or Fock-amplifying property when concatenated with another operator.

We observe that the creation and annihilation operators are Fock-preserving.
As a consequence, any concatenation of such operators yields an operator that is Fock-orthogonal, Fock-amplifying, and Fock-preserving. Thus, following from our Theorems \ref{theorem:single_mode}, \ref{theorem:dual_mode}, \ref{th:concat}, we have shown that any concatenation of photon addition and subtraction on a TMSV always produces a state that is more entangled than the TMSV. 
Finally, we should stress the fact that this result applies to \textit{ideal} photon addition and subtraction.
In the next section, we will address the case of  \textit{realistic} photon addition and subtraction.

\section{Realistic photon addition and subtraction}
\label{sec:realistic}

\begin{figure}[t]
\includegraphics[width=0.95\linewidth]{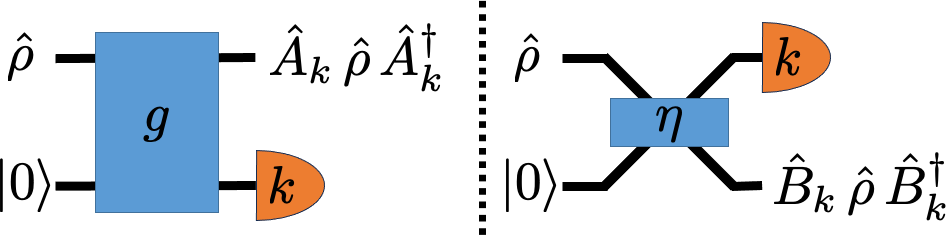}
\caption{
Realistic schemes of photon addition and subtraction.
On the left (resp. right), a two-mode squeezer with gain $g$ (resp. beam-splitter with transmittance $\eta$) acts on an input $\hat{\rho}$ and a vacuum environment $\ket{0}$; then, conditionally on the measurement of $k$ photons at the environment output, the (non-normalized) resulting state is $\hat{A}_k\hat{\rho}\hat{A}^\dagger_k$ (resp. $\hat{B}_k\hat{\rho}\hat{B}^\dagger_k$). 
The expression of the Kraus operators $\hat{A}_k$ and $\hat{B}_k$ is given in Eq. \eqref{eq:krauss_op}.
}
\label{fig:krauss_op}
\end{figure}

The schemes \ref{fig:schemes}(a) and \ref{fig:schemes}(b) presented in Sec. \ref{sec:schemes} use \textit{ideal} versions of photon addition and subtraction.
Indeed, $\hat{a}^k$ and $\hat{a}^{\dagger k}$ are unbounded operators that cannot be exactly implemented in a physical setup.
In practice, photon addition (resp. subtraction) is usually performed using a quantum-limited amplifier (resp. pure-loss channel) followed by post-selection (see Fig. \ref{fig:krauss_op}).
We denote the Kraus operators of the quantum-limited amplifier (with gain $g$) and pure-loss channel (with transmittance $\eta$) as $\hat{A}_k$ and $\hat{B}_k$, respectively \cite{Ivan2011-pm, Sabapathy2017-fl}:
\begin{align}
\begin{split}
    \hat{A}_k
    &=
    \sqrt{\frac{(g-1)^k}{g\, k!}}\sqrt{g}^{\,-\hat{n}}\hat{a}^{\dagger k},
    \\
    \hat{B}_k
    &=
    \sqrt{\frac{(1-\eta)^k}{k!}}\sqrt{\eta}^{\hat{n}}\hat{a}^k.
\end{split}
\label{eq:krauss_op}
\end{align}
Ideal photon addition and subtraction correspond to the limiting cases of $g\rightarrow 1$ and $\eta\rightarrow 1$ (and probability of success going to zero for $k>0$).
At this point, it is interesting to notice that realistic photon addition and subtraction acting on a thermal state produce two states with same eigenvalues, when their efficiencies are related as $\eta\cdot g=1$.
More precisely, the following relation holds:
\begin{align}
    \frac{\hat{A}_k\hat{\tau}\hat{A}^\dagger_k}{\Tr[\hat{A}_k\hat{\tau}\hat{A}^\dagger_k]}
    \equiv
    \frac{\hat{B}_k\hat{\tau}\hat{B}^\dagger_k}{\Tr[\hat{B}_k\hat{\tau}\hat{B}^\dagger_k]},
    \label{eq:equivalence_real_addsub}
\end{align}
under the condition that $\eta\cdot g=1$.
In other terms, this means that realistic photon addition and subtraction are equally good at enhancing entanglement on a TMSV.

It is natural to look whether the realistic operators of Eq. \eqref{eq:krauss_op} fall into the scope of Theorems \ref{theorem:single_mode} or \ref{theorem:dual_mode}.
To verify that, we need to check whether these operators are Fock-orthogonal and Fock-amplifying.
It is easily seen that any of the $\hat{A}_k$ and $\hat{B}_k$ is Fock-orthogonal.
However, none of them is Fock-amplifying as we have 
\begin{align}
\begin{split}
    \Vert\hat{A}_k\ket{n}\Vert\leq\Vert\hat{A}_k\ket{n+1}\Vert
    \quad\Leftrightarrow\quad 
    n\leq\frac{k}{g-1},
    \\
    \Vert\hat{B}_k\ket{n}\Vert\leq\Vert\hat{B}_k\ket{n+1}\Vert
    \quad\Leftrightarrow\quad 
    n\leq\frac{k}{1-\eta}.
    \label{eq:realistic_not_fockamp}
\end{split}
\end{align}
Thus, for $g>1$ or $\eta<1$, there is a threshold on $n$ above which the magnitude of the amplitudes ceases to increase with $n$.
Therefore, Theorems \ref{theorem:single_mode} and \ref{theorem:dual_mode} do not apply to these realistic operators.

\begin{figure}
\includegraphics[width=1\linewidth]{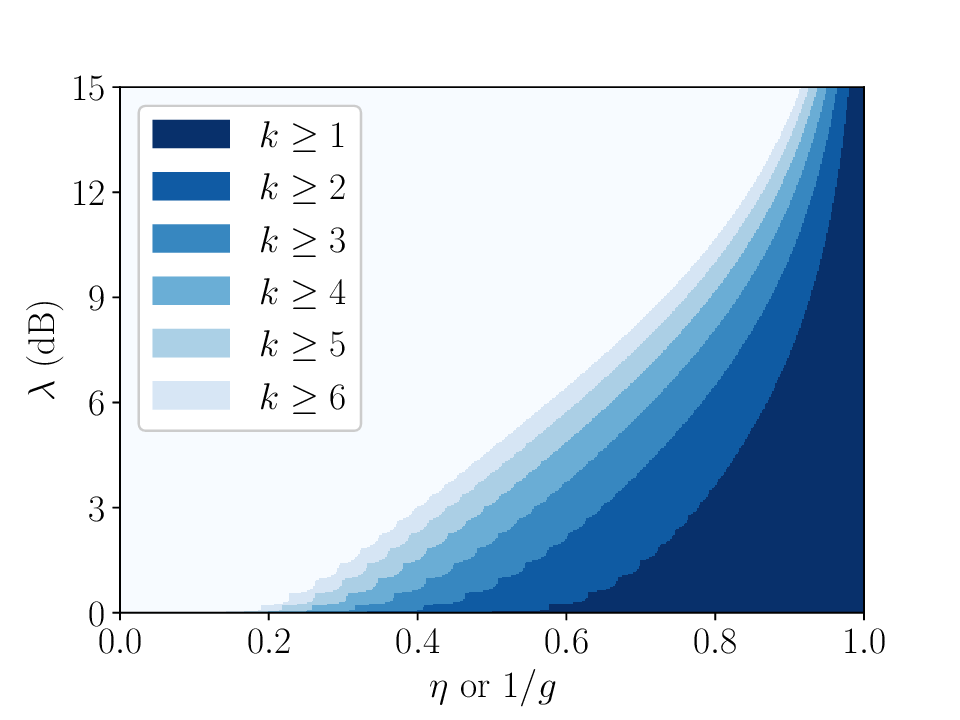}
\caption{
Each point on this graph corresponds to a couple $(\eta,\lambda)$. 
The parameter $\lambda$ defines a thermal state $\hat{\tau}_\lambda$, the parameter $\eta$ defines a set of realistic photon subtraction Kraus operators $\lbrace\hat{B}_k\rbrace$ (see Eq. \eqref{eq:krauss_op}).
We then compare $\hat{\tau}_\lambda$ to $\hat{\sigma}_k=\hat{B}_k\hat{\tau}_\lambda\hat{B}^\dagger_k/\mathrm{Tr}[\hat{B}_k\hat{\tau}_\lambda\hat{B}^\dagger_k]$ for $k\in\lbrace 1,...,6\rbrace$.
The color of the point $(\eta,\lambda)$ is related to the values of $k$ above which entanglement enhancement is successful ($\hat{\sigma}_k\prec\hat{\tau}_\lambda$).
Note that we get identical results for realistic photon addition, since it produces equivalent output as realistic photon subtraction when $g=1/\eta$.
For ideal photon addition ($g=1$) or ideal photon subtraction ($\eta=1$), the majorization relation always holds, as expected from Theorem \ref{theorem:single_mode}.
Note that the squeezing expressed in dB is related to $\lambda$ as: $\lambda[\mathrm{dB}]=(20/\ln10)\tanh^{-1}(\sqrt{\lambda})$.
}
\label{fig:num_addsub}
\end{figure}

Remember now that Theorems \ref{theorem:single_mode} and \ref{theorem:dual_mode} provide us with a \textit{sufficient} condition for majorization, so that the condition not being fullfilled does not imply that there is no majorization relation. Indeed, we observe from numerical simulations that, in some realistic regime ($\eta<1$ or $g>1$), the majorization relation $\hat{\sigma}\prec\hat{\tau}$ does hold (where $\hat{\sigma}=\hat{A}_k\hat{\tau}\hat{A}^\dagger_k/\mathrm{Tr}[\hat{A}_k\hat{\tau}\hat{A}^\dagger_k]\equiv\hat{B}_k\hat{\tau}\hat{B}^\dagger_k/\mathrm{Tr}[\hat{B}_k\hat{\tau}\hat{B}^\dagger_k]$).
This is illustrated on Fig. \ref{fig:num_addsub}. We observe that for a given $\lambda$ and $\eta=1/g$, there is a minimum number of photons that must be added or subtracted in order to achieve entanglement enhancement as witnessed by a majorization relation. For a fixed $\lambda$, this minimum number of photons gets larger when we move far from ideal photon addition/subtraction (\textit{i.e.}, when $\eta=1/g$ gets much smaller than 1). This increase of the minimum number of photons that must be added or subtracted is even faster when $\lambda$ is larger, which means that entanglement enhancement is more sensitive to the non-ideality of photon addition/subtraction in the high-squeezing regime; this makes sense in view of Eq. \eqref{eq:realistic_not_fockamp} together with the fact that the photon number distribution is wider for a TMSV state with higher squeezing.

Note that for reference, we also plotted in Appendix \ref{apd:entropy_of_entanglement} the region of $(\eta,\lambda)$ where the entropy of entanglement increases after photon subtraction, see Fig. \ref{fig:entropy_of_entanglement}.
Comparing with Fig. \ref{fig:num_addsub}, it is apparent that the region where $\hat{\sigma}\prec\hat{\tau}$ is strictly included in the region where $S(\hat{\sigma})\geq S(\hat{\tau})$, as expected from the fact that a majorization relation implies an inequality on entropies but does not assume asymptotic regularization (hence, it is more strict than a simple comparison of entropies).

\subsection*{Approximate majorization}

Let us get a closer look on these realistic operators and see whether we can gain some insights on the existence of a majorization relation.
Interestingly, the operators $\hat{A}_k$ and $\hat{B}_k$ are Fock-orthogonal, so that it is possible to build a matrix $\mathbf{M}$ such that $\bm{\sigma}=\mathbf{M}\bm{\tau}$ (as we did in the first step of Theorem \ref{theorem:single_mode}'s proof).
The matrix $\mathbf{M}$ has the structure of \eqref{eq:structure_columnstocastic} and is defined with respect to a vector $\mathbf{m}$ with components $m_n=\lambda^n(\abs{\varphi_n}^2-\abs{\varphi_{n-1}}^2)/\mathcal{N}$, where $\abs{\varphi_n}=\Vert\hat{A}_k\ket{n}\Vert$ or $\Vert\hat{B}_k\ket{n}\Vert$.
When the index $n$ becomes higher than the threshold of Eq. \eqref{eq:realistic_not_fockamp}, the components $m_n$ become strictly negative, so that the matrix $\mathbf{M}$ is not column-stochastic.
However, observe that the factor $\lambda^n$ becomes close to zero as $n$ grows, so that the negative entries of the matrix $\mathbf{M}$ may all be very close to zero.
This means that, even though $\mathbf{M}$ possesses negative entries, it may actually be \textit{close} to being column-stochastic. Remember also that, in accordance with Fig.~\ref{fig:num_addsub},  another matrix $\mathbf{D}$ connecting  $\bm{\tau}$ to $\bm{\sigma}$  may very well exist that is exactly column-stochastic in case a majorization relation holds, but its structure must then slightly differ from that of Eq. \eqref{eq:structure_columnstocastic}.

With this in mind, let us take a step back and take inspiration from \cite{Horodecki2018-sj}.
Consider that we have two probability vectors 
$\bm{\sigma},\bm{\tau}$ related through a matrix $\mathbf{M}$ as $\bm{\sigma}=\mathbf{M}\bm{\tau}$, where $\mathbf{M}$ has the structure of Eq. \eqref{eq:structure_columnstocastic}, \textit{i.e.}, it is a lower-triangular circulant matrix such that $M_{ij}=m_{i-j}$ for a vector $\mathbf{m}\in\mathbb{R}^{\mathbb{N}}$ (with $m_{n}=0\ \forall n<0$).
Observe that when the vector $\mathbf{m}$ is a probability vector ($m_n\geq 0$, $\forall n$, and $\sum_n m_n=1$), the matrix $\mathbf{M}$ is column-stochastic.
However, as soon as $\mathbf{m}$ possesses one strictly negative component, the matrix $\mathbf{M}$ is \textit{not} column-stochastic.
If that happens, we are unable to conclude that $\bm{\sigma}\prec\bm{\tau}$, whereas it may or may not be the case.
In what follows, we are going to prove a theorem which upper bounds the distance between the vector $\bm{\sigma}$ and another vector $\mathbf{s}$ which is such that $\mathbf{s}\prec\bm{\tau}$ holds for sure. 

\begin{theorem}[Approximate majorization]
Let $\bm{\sigma},\bm{\tau}\in\mathbb{R}^{\mathbb{N}}$ be probability vectors.
Let $\bm{\sigma}=\mathbf{M}\bm{\tau}$, where $\mathbf{M}\in\mathbb{R}^{\mathbb{N}\times\mathbb{N}}$ is such that $M_{ij}=m_{i-j}$ for a vector $\mathbf{m}\in\mathbb{R}^{\mathbb{N}}$ (with $m_n=0$, $\forall n<0$). 
Let $\nu$ be the absolute sum of the negative components of $\mathbf{m}$, \textit{i.e.}, $\nu=\sum_{n\,:\,m_n<0}\abs{m_n}$.
Then, there exists a vector $\mathbf{s}\prec\bm{\tau}$ such that $\delta(\bm{\sigma},\mathbf{s})\leq\nu$, where  $\delta(\cdot,\cdot)$ stands for the total variation distance.
\label{theorem:majorization_tvd}
\end{theorem}

\begin{proof}
Let us define the vector $\bm{\varepsilon}$ with components $\varepsilon_n\vcentcolon=-\min(m_n, 0)$, and the matrix $\mathbf{E}$ with entries $E_{ij}\vcentcolon=-\min(M_{ij},0)$.
The vector $\bm{\varepsilon}$ only contains the absolute value of the negative components of $\mathbf{m}$, and is filled with zeros at the positive components of $\mathbf{m}$.
The matrix $\mathbf{E}$ is built similarly with respect to $\mathbf{M}$.
Using this, we can then define the matrix $\mathbf{D}=(\mathbf{M}+\mathbf{E})/\alpha$, where $\alpha$ is a normalization constant defined as $\alpha=1+\sum_n\varepsilon_n$.
Observe that the matrix $\mathbf{D}$ is column-stochastic by definition: it has only non-negative entries, the sum of each column is $1$, the sum of each row is less or equal to $1$.
Starting from $\bm{\sigma}=\mathbf{M}\bm{\tau}$, we may write $\bm{\sigma}=(\alpha\mathbf{D}-\mathbf{E})\bm{\tau}$ or equivalently $\bm{\sigma}=\alpha\mathbf{D}\bm{\tau}-\mathbf{E}\bm{\tau}$.
Introducing then the vector $\mathbf{s}\vcentcolon=\mathbf{D}\bm{\tau}$, we have $\bm{\sigma}=\alpha\mathbf{s}-\mathbf{E}\bm{\tau}$.
Now, from the column-stochasticity of $\mathbf{D}$, observe that the majorization relation $\mathbf{s}\prec\bm{\tau}$ holds.

The next step of our reasoning is to evaluate the total variation distance (TVD) between $\bm{\sigma}$ and $\mathbf{s}$.
The TVD is defined from the $\ell_1$ norm, which for vectors is $\Vert\mathbf{p}\Vert=\sum_n\abs{p_n}$, and for matrices is $\Vert\mathbf{M}\Vert=\max_{j}\sum_i\abs{M_{ij}}$.
The TVD between $\bm{\sigma}$ and $\mathbf{s}$ is then $\delta(\bm{\sigma},\mathbf{s})\vcentcolon=(1/2)\Vert\bm{\sigma}-\mathbf{s}\Vert$.
Observe that $\Vert\mathbf{E}\Vert=\Vert\bm{\varepsilon}\Vert$ and $\alpha=1+\Vert\bm{\varepsilon}\Vert$.
Note that $\Vert\bm{\varepsilon}\Vert$ is the absolute value of the sum of all the negative entries of $\mathbf{m}$ (so that $\Vert\bm{\varepsilon}\Vert=\nu$ in the statement of Theorem \ref{theorem:majorization_tvd}).
We can then upper bound $\delta(\bm{\sigma},\mathbf{s})$ as follows:
\begin{align}
\begin{split}
    \delta(\bm{\sigma},\mathbf{s})&=
    \frac{1}{2}
    \Vert\bm{\sigma}-\mathbf{s}\Vert
    \\&=
    \frac{1}{2}
    \Vert
    (\alpha-1)\mathbf{s}-\mathbf{E}\bm{\tau}
    \Vert
    \\
    &\leq
    \frac{1}{2}
    \abs{\alpha-1}\cdot\Vert\mathbf{s}\Vert
    +\frac{1}{2}
    \Vert\mathbf{E}\bm{\tau}\Vert
    \\&\leq
    \frac{1}{2}
    \Vert\bm{\varepsilon}\Vert
    +
    \frac{1}{2}
    \Vert\mathbf{E}\Vert
    \\&=
    \Vert\bm{\varepsilon}\Vert
\end{split}
\end{align}
The first inequality comes from the triangle inequality while the second one comes from the matrix norm inequality for vectors ($\Vert\mathbf{E}\bm{\tau}\Vert\leq\Vert\mathbf{E}\Vert\cdot\Vert\bm{\tau}\Vert$).
We also have used $\Vert\mathbf{s}\Vert=\Vert\bm{\tau}\Vert=1$, and $\Vert\mathbf{E}\Vert=\Vert\bm{\varepsilon}\Vert$.
This concludes our proof, as we have shown that $\mathbf{s}\prec\bm{\tau}$ and $\delta(\bm{\sigma},\mathbf{s})\leq\Vert\bm{\varepsilon}\Vert=\nu$.
\end{proof}
As a conclusion, the value of $\Vert\bm{\varepsilon}\Vert$ gives us a good indicator on the maximum distance between $\bm{\sigma}$ and $\mathbf{s}$.
The smaller the value of $\Vert\bm{\varepsilon}\Vert$, the closer to $\hat{\sigma}$ it is guaranteed that there  exists a state for which the majorization relation holds.
As implied by Fig.~\ref{fig:num_addsub}, remember that a non-zero value of the TVD $\delta$ does not automatically imply that the majorization relation between $\bm{\sigma}$ and $\bm{\tau}$ does not hold (but we can only check numerically whether majorization holds or not). In practice, however, it seems that majorization holds provided $\delta$ is sufficiently small (see Fig. \ref{fig:major_dist}).

The upper bound on the TVD we have derived may now be applied to uniform continuity bounds for various functionals.
The particular case of the Shannon entropy for infinite-dimensional vectors was studied in Ref. \cite{Becker2023-rh}.
Recall that the Shannon entropy of a probability vector $\mathbf{p}$ is defined as $H(\mathbf{p})\vcentcolon=-\sum_n p_n\ln p_n$, and that the von Neumann entropy of a quantum state $\hat{\rho}$ is the Shannon entropy of its eigenvalues, \textit{i.e.}, $S(\hat{\rho})\vcentcolon=H(\bm{\lambda}(\hat{\rho}))$.
Introducing the binary entropy $h_2(x)\vcentcolon=-x\ln x-(1-x)\ln(1-x)$ and the mean photon number of a probability vector $\mathbf{p}$ as $N_\mathbf{p}=\sum_n n\, p_n$, it is shown that
\begin{align}
    \abs{H(\bm{\sigma})-H(\mathbf{s})}
    &\leq
    h_2\big(\delta(\bm{\sigma},\mathbf{s})\big)+Nh_2\big(\delta(\bm{\sigma},\mathbf{s})/N\big)
    \nonumber
    \\[0.8em]
    &\leq
    h_2(\Vert\bm{\varepsilon}\Vert)+Nh_2(\Vert\bm{\varepsilon}\Vert/N)
    \label{eq:continuity_bound}
\end{align}
where $N=\max(N_{\bm{\sigma}},N_{\mathbf{s}})$.
The second inequality of Eq. \eqref{eq:continuity_bound} holds as soon as $\Vert\bm{\varepsilon}\Vert\leq 1/2$, since $h_2$ is non-decreasing over $[0,1/2]$.
Eq. \eqref{eq:continuity_bound} is useful because it sets an upper bound to the quantity $H(\bm{\sigma})-H(\mathbf{s})$, which is itself an upper bound on $H(\bm{\sigma})-H(\bm{\tau})$ (since $\mathbf{s}\prec\bm{\tau}$).
In practice, the quantity $H(\bm{\sigma})=S(\hat{\sigma})$ is known as the entropy of entanglement of the bipartite pure state $\ket{\Phi}$ (such that $\hat{\sigma}=\mathrm{Tr}_2[\ket{\Phi}\bra{\Phi}]$), and is the most common measure of entanglement for pure states.

\subsection*{Application to realistic single-photon addition}

Let us now apply Theorem \ref{theorem:majorization_tvd} to realistic photon addition.
Note that our results will extend seamlessly to realistic photon subtraction, since they produce equivalent states
[when $g\cdot\eta=1$, see Eq. \eqref{eq:equivalence_real_addsub}].
In order to make that equivalence more obvious, we introduce the parameter $\mu=\lambda/g$, which should simply be set to $\mu=\eta\lambda$ in the case of realistic photon subtraction.
Observe that $\mu\in[0,1]$ and $\mu\leq\lambda$.

From now on, we define the vector $\mathbf{m}$ as $m_n=\lambda^n(\abs{\varphi_n}^2-\abs{\varphi_{n-1}}^2)/\mathcal{N}$, where $\abs{\varphi_n}=\Vert\hat{A}_k\ket{n}\Vert$ (with $\abs{\varphi_{n}}=0\ \forall n<0$).
The normalization constant is $\mathcal{N}=\sum_n \lambda^n(\abs{\varphi_n}^2-\abs{\varphi_{n-1}}^2)$.
The components of $\mathbf{m}$ can be computed to the following:
\begin{align}
    m_n
    &=
    \frac{(1-\mu)^{k+1}\mu^{n-1}}{1-\lambda}
    \frac{(k+n-1)!}{k!n!}
    \big(k\mu+n\mu-n\lambda\big)
    ,
    \nonumber
\end{align}
which in turn defines the matrix $\mathbf{M}$ with components $M_{ij}=m_{i-j}$.
The matrix $\mathbf{M}$ is such that $\bm{\sigma}=\mathbf{M}\bm{\tau}$, where $\bm{\tau}$ is the vector of eigenvalues of the thermal state $\hat{\tau}$ and $\bm{\sigma}$ is the vector of eigenvalues of the realistic photon-added thermal state $\hat{\sigma}=\hat{A}_k\hat{\tau}\hat{A}^\dagger_k/\mathrm{Tr}[\hat{A}_k\hat{\tau}\hat{A}^\dagger_k]$.
We define the vector $\bm{\varepsilon}$ as $\varepsilon_n=-\min(m_n, 0)$.
Observe that $m_n$ is negative if and only if $n\geq k/(g-1)=\mu k/(\lambda-\mu)$.
In order to compute the TVD between $\bm{\sigma}$ and another state $\mathbf{s}$ such that $\mathbf{s}\prec\bm{\tau}$, we need to sum all the negative entries of $\mathbf{m}$.
We define the partial sum of the last components of $\mathbf{m}$ starting from index $p$ as $\Sigma(p)$.
This evaluates to the following:
\begin{align}
    \Sigma(p)
    \vcentcolon=&
    \sum\limits_{n=p}^{\infty}m_n=
    \mu^p(1-\mu)^k(k+p-1)!
    \label{eq:partial_summation}
    \\
    &\times\left[
    \frac{\lambda-\mu}{(\lambda-1)\mu k!(p-1)!}
    +
    \frac{F(1,k+p;p+1;\mu)}{(k-1)!}
    \right]
    \nonumber
\end{align}
where $F(a,b;c;z)$ is the regularized hypergeometric distribution.
It is remarkable to obtain a closed expression for the summation here (this works for any $k\in\mathbb{N}\setminus\lbrace 0\rbrace$).

At this point, we will focus on realistic single-photon addition, \textit{i.e.}, set $k=1$.
Eq. \eqref{eq:partial_summation} then becomes:
\begin{align}
    \Sigma(p)\Big\vert_{k=1}
    =
    \mu^p
    \left[
    1-\frac{(\lambda-\mu)(1-\mu)}{(1-\lambda)\mu}\;p
    \right].
    \label{eq:partial_summation_k1}
\end{align}
The component $m_n$ becomes negative as soon as $n\geq 1/(g-1)$.
If $1/(g-1)\in\mathbb{N}$, we just need to insert $p=1/(g-1)=\mu/(\lambda-\mu)$ in $\Sigma(p)$ to get the exact solution for $\Vert\bm{\varepsilon}\Vert=\Sigma(p)$, which will read as follows:
\begin{align}
    \Vert\bm{\varepsilon}\Vert
    =
    \exp_{\mu}\left(\frac{\mu}{\lambda-\mu}\right)
    \frac{\lambda-\mu}{1-\lambda},
\end{align}
where $\exp_\mu(x)\vcentcolon=\mu^x$.
Keep it mind that the above relation only holds when $\mu/(\lambda-\mu)\in\mathbb{N}$, and observe that this approximately holds as soon as $g$ or $\eta$ is close to $1$.

In general, however, $\mu/(\lambda-\mu)$ is not an integer and we will upper bound $\Vert\bm{\varepsilon}\Vert$.
To do so, simply note that the sum of all the negative components of $\mathbf{m}$ is greater or equal to the minimum of $\Sigma(p)$.
Let us define $p^\ast$ as the argument of the minimum of $\Sigma(p)$, so that $\min_p\Sigma(p)=\Sigma(p^\ast)$ (note that $p^\ast\in\mathbb{R}$).
The value of $p^\ast$ can be computed to the following:
\begin{align}
    p^\ast
    &=\frac{\mu(1-\lambda)}{(1-\mu)(\lambda-\mu)}
    -\frac{1}{\ln\mu}.
\end{align}
We then have $\Vert\bm{\varepsilon}\Vert\leq\abs{\Sigma(p^\ast)}$, which yields:
\begin{align}
\begin{split}
    \Vert\bm{\varepsilon}\Vert
    \;\leq\;&
    \exp_{\mu}\left(\frac{\mu(1-\lambda)}{(\lambda-\mu)(1-\mu)}\right)
    \\[0.6em]
    &\times\;
    \frac{1}{e\ln\mu}\;
    \frac{\mu(1-\lambda)}{(1-\mu)(\lambda-\mu)}.
\end{split}
\label{eq:bound_major_dist}
\end{align}
Remember that the above expression also applies to realistic single-photon subtraction if we set $\mu=\eta\lambda$.
The upper bound \eqref{eq:bound_major_dist} holds for every possible value of $\lambda$ and $\mu$ (every gain $g$ or transmittance $\eta$).

\begin{figure}
\includegraphics[width=\linewidth]{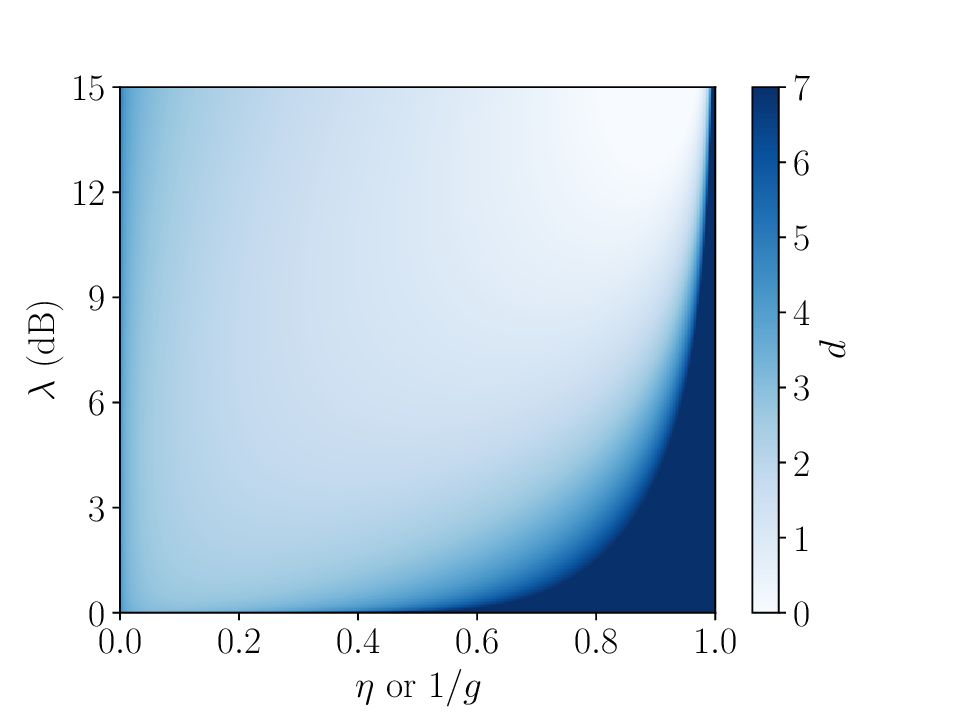}
\caption{
Each couple $(\eta,\lambda)$ defines a thermal state $\hat{\tau}$ and a state $\hat{\sigma}=\hat{A}_1\hat{\tau}\hat{A}^\dagger_1/\mathrm{Tr}[\hat{A}_1\hat{\tau}\hat{A}^\dagger_1]$, where $\hat{A}_1$ is the realistic single photon addition operator (see Eq. \eqref{eq:krauss_op} with $k=1$).
This logarithmic plot associates each point $(\eta,\lambda)$ to a value $d$, such that we can ensure that the state $\hat{\sigma}$ is at a distance of at most $10^{-d}$ (in TVD) from another state that is majorized by $\hat{\tau}$, see Eq. \eqref{eq:bound_major_dist}.
Note that no distinction is made among values above $7$, and among values below $0$.
This figure should be compared with Fig.~\ref{fig:num_addsub} for $k=1$.
}
\label{fig:major_dist}
\end{figure}

We plot the upper bound \eqref{eq:bound_major_dist} in Fig.~\ref{fig:major_dist}.
It is interesting to observe that the region where the relation $\bm{\sigma}\prec\bm{\tau}$ holds (as shown in Fig. \ref{fig:num_addsub} for $k=1$) and the region where we can ensure a sufficiently small TVD (as shown in Fig. \ref{fig:major_dist} for $\delta(\bm{\sigma},\mathbf{s})\lesssim
10^{-7}$) accurately coincide.
It was expected that, for a small enough TVD,  the majorization relation $\bm{\sigma}\prec\bm{\tau}$ would hold (so that the small-TVD region is included in the majorization region), but it is remarkable that the small-TVD region covers most of the majorization region (as soon as $\delta(\bm{\sigma},\mathbf{s})\lesssim 10^{-7}$).
Indeed, there could have existed regions where the majorization relation hold but we are not able to ensure a small TVD; Figs \ref{fig:num_addsub} and \ref{fig:major_dist} show that it is not the case.
This illustrates the practical applicability of our Theorem \ref{theorem:majorization_tvd} even for realistic photon addition and subtraction.

\section{Discussion and conclusion}
\label{sec:concl}

Throughout this paper, we have investigated the enhancement of the entanglement of a TMSV state by using local operations on each of its two mode.
We presented in Sec. \ref{sec:schemes} several schemes based on photon addition and subtraction, and then introduced a more general filtration scheme consisting in applying some local (non-unitary) operator on each of the two modes.

Our main analytical result was presented in Sec. \ref{sec:results}, where we provided a set of sufficient conditions on the filtration operators in order to produce a state that is more entangled than the original TMSV state.
It is remarkable that a set of only two properties is sufficient to guarantee this entanglement enhancement: the filtration operators must be Fock-orthogonal and jointly Fock-amplifying.
It is not surprising that these properties give a particular importance to the Fock basis as it is the natural basis of the TMSV, \textit{i.e.}, the basis appearing in its Schmidt decomposition.
The criterion we used to compare entanglement relied on majorization theory through Nielsen's theorem.
The main ingredient of our proof is the derivation of a column-stochastic matrix that  exploits the intrinsic symmetry of the TMSV and corresponding reduced thermal states, whose eigenvalues obey a geometric distribution.
Our result implies in turn that any concatenation of the (ideal) creation and annihilation operators on the TMSV produce states that are more entangled than the original TMSV.

In Sec. \ref{sec:realistic}, we addressed the case of realistic photon addition and subtraction as realized with a beam splitter or two-mode squeezed supplemented with post-selection. The associated filtration operators are not Fock-amplifying, in contrast to their ideal counterparts.
Thus, as such, these operators are not concerned by our theorems.
Nevertheless, in the case of filtration scheme based on realistic photon addition or subtraction, we were able to set an upper bound on the distance between the state that is actually produced and a state for which the majorization relation provably holds.
To some extent, this upper bound gives us a figure of merit on the efficiency of the scheme.

Another issue that is worth discussing is that the filtration schemes analyzed here are inherently associated with a success probability $\mathcal{P}\leq 1$.
A detailed exploration of the relationship between $\mathcal{P}$ and the amount of entanglement enhancement that is achieved would be  meaningful if we were focusing on a specific measure of entanglement, such as the entanglement entropy, rather than the majorization theoretical approach considered here, which yields a dichotomic criterion (the majorization relation either holds or not).
Nevertheless, we are able to provide some general intuition about the interplay between the Fock-amplifying property and the success probability.
To illustrate this, consider a single-mode filtration scheme employing a Fock-amplifying and Fock-orthogonal operator $\hat{F}$ acting on the Fock basis as $\hat{F}\ket{n}=\varphi_n\ket{\phi_n}$.
In practice, filtration operators are Kraus operators, meaning that $\hat{F}$ belongs to a set $\lbrace\hat{F}_k\rbrace$ such that $\sum_k\hat{F}^\dagger_k\hat{F}_k=\hat{\mathds{1}}$ and $\hat{F}^\dagger_k\hat{F}_k\geq 0$.
This condition implies that $\abs{\varphi_n}\leq 1$.
The probability of success $\mathcal{P}$ of the local filtration scheme is then expressed as
\begin{align}
    \mathcal{P}
    &=
    \big\Vert
    \hat{F}
    \ket{\Psi_\lambda}
    \big\Vert^{2}
    =
    \sum\limits_{n}
    \abs{\varphi_n}^2
    \tau_n.
    \label{eq:success_probability}
\end{align}
Since $\abs{\varphi_n}^2\leq 1$, the above equation implies that $\mathcal{P}\leq 1$, as expected.
From Eq.~\eqref{eq:success_probability}, the success probability can be interpreted as an overlap between the eigenvalues $\tau_n$ of the thermal state and the squared amplitudes $\abs{\varphi_n}^2$ of the Fock-amplifying operator $\hat{F}$.
Now, observe that the eigenvalues $\tau_n=(1-\lambda)\lambda^n$ decrease with $n$, whereas the squared amplitudes $\abs{\varphi_n}^2$ increase with $n$ and eventually reach a plateau at (or below)~1.
The more gradual is this increase of $\abs{\varphi_n}^2$ with $n$, the smaller is the resulting overlap: this reveals a trade-off between the Fock-amplifying character of $\hat{F}$ and the success probability $\mathcal{P}$.
Additionally, as the parameter $\lambda$ increases (higher energy TMSV state), the barycenter of the eigenvalues $\tau_n$ shifts towards higher $n$, leading to an increased success probability $\mathcal{P}$ for a fixed $\hat{F}$. 
Note that all the above reasoning naturally extends to a pair of jointly Fock-amplifying operators used in a dual-mode scheme.

We take advantage of the present discussion to address some further considerations. In our analysis until now, one of the two states whose entanglement we compared was always the TMSV.
However, it is natural to try to compare states produced from different setups, in order to determine whether one entanglement-enhancement setup is more efficient than another one, for example.
The tasks appears to be much harder as we cannot exploit anymore the simple structure of the TMSV.
Nevertheless, in some cases, we could still establish a majorization relation from previously proven results.

In particular, let us consider the case of dual-mode single photon addition/subtraction, see Fig. \ref{fig:schemes}(a) and Eq. \eqref{eq:phi_dualmode_single}.
Here, $\hat{\sigma}^{k,l}$ is the partial trace of a TMSV with $k$ photons added/subtracted on mode 1 and $l$ photons added/subtracted on mode 2 (depending on the sign of $k$ and $l$).
According to Sec.~\ref{sec:results}, we know $\hat{\sigma}^{k,\ell}\prec\hat{\sigma}^{0,0}$.
Ref. \cite{navarrete2012enhancing} proves the following relation (with $k\geq 0$):
\begin{align}
    \hat{a}^{\dagger k}\,\hat{U}_\lambda\ket{0,0}\propto\hat{b}^k\,\hat{U}_\lambda\ket{0,0}
    \propto\hat{U}_\lambda\hat{a}^{\dagger k}\ket{0,0},
    \label{eq:tms_commut_channel}
\end{align}
where the sign $‘‘\propto"$ denotes that the states are equal up to a normalization constant.
From this, we observe that $\hat{\sigma}^{k,0}=\hat{\sigma}^{0,-k}=\Tr_2[\hat{U}_\lambda\left(\ket{k}\bra{k}\otimes\ket{0}\bra{0}\right)\hat{U}_\lambda^\dagger]$ is the output of the Fock state $k$ through a quantum-limited amplifying channel.
This is a bosonic Gaussian channel and thus obeys the majorization ladder for Fock states \cite{Garcia-Patron2012-bn, Gagatsos2013-jy, Van_Herstraeten2023-bw}, which implies the following:
\begin{align}
\begin{split}
    \hat{\sigma}^{k,0}
    &\succ
    \hat{\sigma}^{k+1,0}
    \qquad\quad
    \forall k\in\mathbb{N},
    \\
    \hat{\sigma}^{-k,0}
    &\succ
    \hat{\sigma}^{-k-1,0}
    \qquad\ 
    \forall k\in\mathbb{N}.
\end{split}
\label{eq:major_ladder_bgc}
\end{align}
Eq. \eqref{eq:major_ladder_bgc} is interesting because it allows us to compare different entanglement-enhanced TMSV states among them: we observe that the more photons we add/subtract on one mode (the second mode being untouched), the more the resulting state becomes entangled.
Actually, even if we apply a fixed photon addition (or subtraction) on the second mode, we observe from numerics that the relation $\sigma^{0,l}\succ \sigma^{k,l}$ appears to hold in general, for all $k,l\in\mathbb{N}$ and all $\lambda$.

In the same spirit, it is tempting to look for majorization relations between the states $\hat{\sigma}^{k,k}$ and $\hat{\sigma}^{l,l}$ (for $k,l\in\mathbb{N}$).
Indeed, we have been able to prove particular instances of that relation, namely the cases $\hat{\sigma}^{1,1}\succ\hat{\sigma}^{k,k}$ for $k\in\lbrace 2,...,8\rbrace$ (see Appendix \ref{apd:extra_majorization_relations}) and we conjecture the validity of the aforementioned majorization relation for all higher $k$ (we have not found any numerical counterexample to this conjecture).

In Ref. \cite{navarrete2012enhancing}, it is observed that the entropy of entanglement of the state 
$\hat{\sigma}^{k,k}$ grows monotonically with $k$.
Following this,
another natural majorization chain could be that the state $\hat{\sigma}^{k,k}$ majorizes the state $\hat{\sigma}^{k+1,k+1}$ (with $k\in\mathbb{N}$), but we found instances of $k$ such that there is no majorization relation (for example, we found that $\hat{\sigma}^{8,8}\not\succ\hat{\sigma}^{9,9}$ for $\lambda=0.015$).
Ref. \cite{navarrete2012enhancing} also points out that for a total number of photon additions (resp., photon subtractions) fixed to $2K$, the state yielding the greatest entropy of entanglement is achieved when the photon additions (resp., photon subtractions) are split equally among both modes, \textit{i.e.}, for the state $\sigma^{K,K}$.
This could suggest that a majorization relation holds between the states $\sigma^{K,K}$ and $\sigma^{k,2K-k}$ (for $k\in\lbrace 0,..., 2K\rbrace$); however we found from numerics that the two states are sometimes incomparable.

Finally, we would like to make an observation about Eq. \eqref{eq:tms_commut_channel}.
The quantum-limited amplifier belongs to the family of phase-insensitive bosonic Gaussian channels (BGCs), which has been proven to obey a fundamental majorization relation: the output associated to vacuum majorizes any other output \cite{Mari2014-mn, giovannetti2015majorization}.
At first sight, it may seem possible to use Eq. \eqref{eq:tms_commut_channel} in order to commute an arbitrary number of creation/annihilation operator with the TMS unitary.
From the fundamental majorization relation at the output of BGCs, this would be an easy proof that any concatenation of creation or annihilation operators produces a state more entangled than the TMSV.
However, that reasoning would be flawed because the relation $\hat{a}^{\dagger k}\hat{U}_\lambda\ket{0,0}\propto\hat{U}_\lambda\hat{a}^{\dagger k}\ket{0,0}$ does not generalize to arbitrary states (\textit{i.e.}, $\hat{a}^{\dagger k}\hat{U}_\lambda\ket{m,n}\not\propto\hat{U}_\lambda\hat{a}^{\dagger k}\ket{m,n}$ for arbitrary $m,n$).
It is nevertheless true that any concatenation of creation and annihilation operators enhance the entanglement of a TMSV (as we have shown in Sec. \ref{sec:results}).

\acknowledgements

NJC is grateful to the James C. Wyant College of Optical Sciences for hospitality during his sabbatical leave in the autumn 2022, when this work was completed. ZVH, CNG, and SG acknowledge support from the Department of Energy project on continuous variable quantum networking, under an Oak Ridge National Laboratory (ORNL) contract. CNG acknowledges funding support from the National Science Foundation, FET, Award No. 2122337. ZVH received partial funding support from the ARO Quantum Network Science MURI project. NJC acknowledges support from the European Union and the Fonds de la Recherche Scientifique – FNRS  under project ShoQC within ERA-NET Cofund in Quantum Technologies (QuantERA) program.

\bibliography{bib}

\newpage
\section*{Appendices}
\appendix

\section{Single addition vs. subtraction identity}
\label{apd:TMS-commutation-identity}

\begin{figure}[t]
\includegraphics[width=1\linewidth]{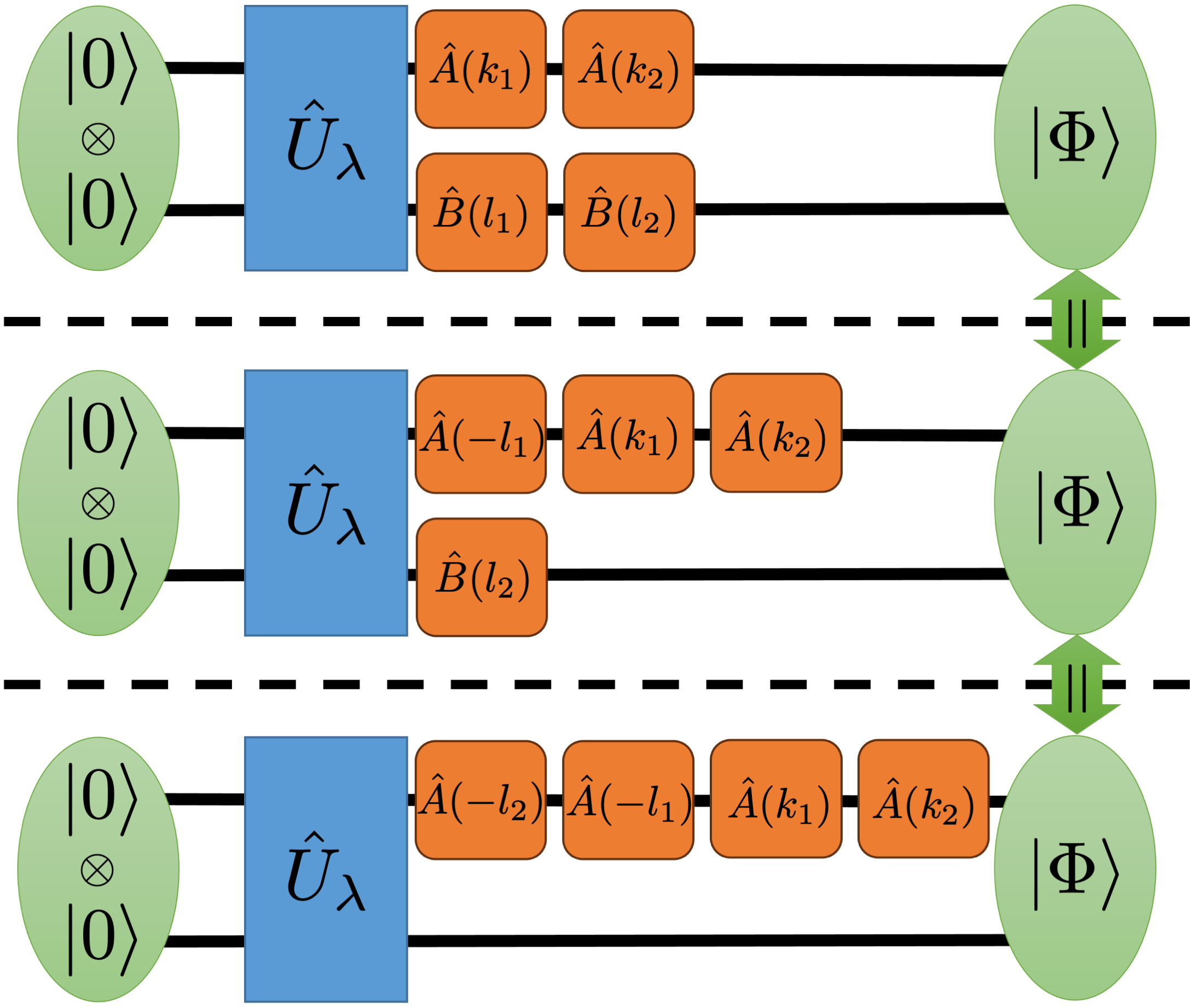}
\caption{
 The three above schemes yield the same pure state $\ket{\Phi}$ (after normalization).
 Each (orange) round-cornered box is a photon addition/subtraction ($k_i,l_i\in\mathbb{Z}$). 
Using the TMS commutation identity \eqref{eq:add-sub-id}, we can go from one scheme to the next/previous one.
With multiple uses of the identity, all the photon additions/subtractions can be brought to mode 1 (or mode 2).
This naturally generalizes to an arbitrary number of photon additions/subtractions.
}
\label{fig:add_sub}
\end{figure}

The relation \begin{align} \hat{b}\ket{\Psi_\lambda}=\sqrt{\lambda}\, \hat{a}^{\dagger}\ket{\Psi_\lambda},
    \label{eq:add-sub-id}
    \end{align}
originates from a commutation relation between the creation/annihilation operators and the TMS when it acts on vacuum (see Ref. \cite{navarrete2012enhancing}).
Eq. \eqref{eq:add-sub-id} is a very powerful identity in our concern.
First, notice from the mode exchange invariance of the TMSV state that it implies the corollary $\hat{a}\ket{\Psi_\lambda}=\sqrt{\lambda}\, \hat{b}^\dagger\ket{\Psi_\lambda}$.
 Now, observe that using Eq. \eqref{eq:add-sub-id} (and its corollary) an arbitrarily number of times, it is always possible to bring any concatenation of creation/annihilation operators acting on both modes to a concatenation of creation/annihilation operators acting on a single mode (with no operation applied to the other mode).
 This is illustrated in Fig. \ref{fig:add_sub}.

\section{Proof of Theorem~\ref{th:concat} (concatenability)}
\label{apd:proofs}

\textbf{Theorem~\ref{th:concat}(a)}:
If $\hat{F},\hat{G}$ are Fock-preserving, then $\hat{G}\hat{F}$ is Fock-preserving.

\begin{proof}
When $\hat{F}$ and $\hat{G}$ are Fock-preserving, we can use Eq. \eqref{eq:fock-concatenable} to write $\hat{F}\ket{n}=\varphi_n\ket{m_n}$ and $\hat{G}\ket{n}=\gamma_n\ket{p_n}$, where $\mathbf{m}$ and $\mathbf{p}$ are vectors with non-decreasing integer components.
Using this, we compute the action of $\hat{G}\hat{F}$ onto Fock states as:
\begin{align}
    \hat{G}\hat{F}\ket{n}
    &=
    \hat{G}\,\varphi_{n}\ket{m_n},
    \nonumber
    \\
    &=
    \gamma_{m_n}\varphi_n
    \ket{p_{m_n}},
    \nonumber
    \\
    &=
    \gamma_{m_n}\varphi_n
    \ket{q_n},
    \nonumber
\end{align}
where we have defined the vector $\mathbf{q}$ with components $q_n\vcentcolon=p_{m_n}$.
Observe that both $\mathbf{m}$ and $\mathbf{p}$ are vectors with non-decreasing integer components.
Since a non-decreasing function of a non-decreasing function is non-decreasing, it follows that the vector $\mathbf{q}$ also has non-decreasing integer components.
Thus, $\hat{F}\hat{G}$ is Fock-preserving.
\end{proof}

\textbf{Theorem~\ref{th:concat}(b)}:
If $\hat{F}$ is Fock-preserving and $\hat{F},\hat{G}$ are Fock-orthogonal, then $\hat{G}\hat{F}$ is Fock-orthogonal.

\begin{proof}
The operator $\hat{F}$ is Fock-preserving and Fock-orthogonal, so that it acts on the Fock basis as $\hat{F}\ket{n}=\varphi_n\ket{m_n}$, where the components of the vector $\mathbf{m}$ are strictly increasing (since $\hat{F}$ is Fock-orthogonal).
The operator $\hat{G}$ is Fock-orthogonal, so that it acts on the Fock-basis as $\hat{G}\ket{n}=\gamma_n\ket{\psi_n}$, where $\lbrace\ket{\psi_n}\rbrace$ is an orthonormal set.
Then, we compute the action of $\hat{G}\hat{F}$ onto Fock states as:
\begin{align}
    \hat{G}\hat{F}\ket{n}
    &=
    \hat{G}\,\varphi_n\ket{m_n},
    \nonumber
    \\
    &=
    \gamma_{m_n}\varphi_{n}
    \ket{\psi_{m_n}}.
    \nonumber
\end{align}
From the orthonormality of the set $\lbrace\ket{\psi_n}\rbrace$ and the fact that the components $m_n$ are strictly increasing, it follows that $\lbrace\ket{\psi_{m_n}}\rbrace$ is also orthonormal.
As a consequence, the operator $\hat{G}\hat{F}$ is Fock-orthogonal.
\end{proof}

\textbf{Theorem~\ref{th:concat}(c)}:
If $\hat{F}$ is Fock-preserving and $\hat{F},\hat{G}$ are Fock-amplifying, then $\hat{G}\hat{F}$ is Fock-amplifying.
\begin{proof}
The operator $\hat{F}$ is Fock-preserving and Fock-amplifying, so that it acts on the Fock basis as $\hat{F}\ket{n}=\varphi_n\ket{m_n}$, where $\abs{\varphi_n}\leq\abs{\varphi_{n+1}}$ and $m_n\leq m_{n+1}$.
The operator $\hat{G}$ is Fock-amplifying, so that it acts on the Fock basis as $\hat{G}\ket{n}=\gamma_n\ket{\psi_n}$, where $\abs{\gamma_n}\leq\abs{\gamma_{n+1}}$.
Then, we compute the action of $\hat{G}\hat{F}$ onto Fock states as:
\begin{align}
    \hat{G}\hat{F}\ket{n}
    &=
    \hat{G}\,\varphi_n\ket{m_n},
    \nonumber
    \\
    &=
    \gamma_{m_n}\varphi_n
    \ket{\psi_{m_n}}.
    \nonumber
\end{align}
Since $m_n\leq m_{n+1}$ and $\abs{\gamma_n}\leq\abs{\gamma_{n+1}}$, it follows that $\gamma_{m_n}\leq\gamma_{m_{n+1}}$.
Then, since $\abs{\varphi_n}\leq\abs{\varphi_{n+1}}$, we have $\abs{\gamma_{m_n}\varphi_n}\leq\abs{\gamma_{m_{n+1}}\varphi_{n+1}}$, so that $\hat{G}\hat{F}$ is Fock-amplifying.
\end{proof}

\section{Entropy of entanglement for realistic photon addition and subtraction}
\label{apd:entropy_of_entanglement}

\begin{figure}
\includegraphics[width=1\linewidth]{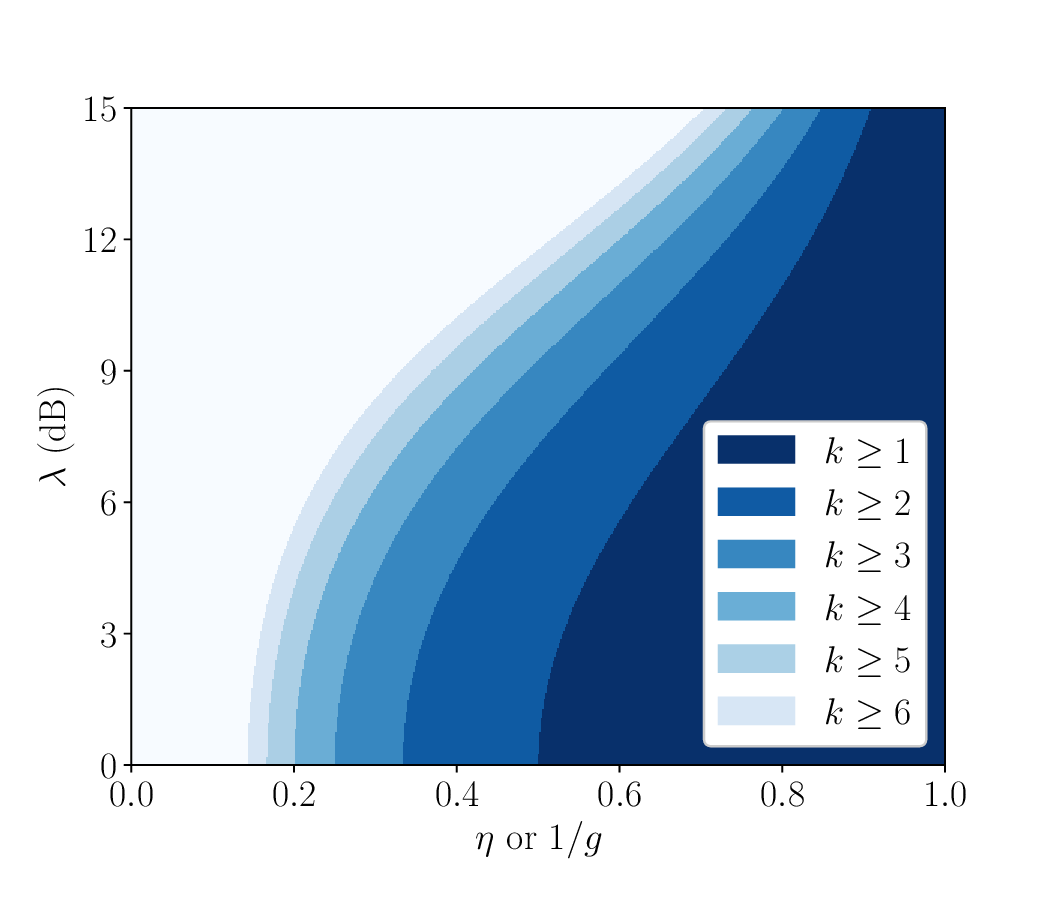}
\caption{
Each point on this graph corresponds to a couple $(\eta,\lambda)$. 
The parameter $\lambda$ defines a thermal state $\hat{\tau}_\lambda$, the parameter $\eta$ defines a set of realistic photon subtraction Kraus operators $\lbrace\hat{B}_k\rbrace$ (see Eq. \eqref{eq:krauss_op}).
We then compute the von Neumann entropy of $\hat{\tau}_\lambda$ and of $\hat{\sigma}_k=\hat{B}_k\hat{\tau}_\lambda\hat{B}^\dagger_k/\mathrm{Tr}[\hat{B}_k\hat{\tau}_\lambda\hat{B}^\dagger_k]$ for $k\in\lbrace 1,...,6\rbrace$.
The color of the point $(\eta,\lambda)$ is related to the values of $k$ above which the entropy of entanglement has been increased after photon subtraction, \textit{i.e.}, $S(\hat{\sigma}_k)\geq S(\hat{\tau}_\lambda)$.
Note that the above graph only contains information about the sign of $S(\hat{\sigma}_k)-S(\hat{\tau}_\lambda)$, not its magnitude.
}
\label{fig:entropy_of_entanglement}
\end{figure}

In Section \ref{sec:realistic}, we interested ourselves to the case of realistic photon addition and subtraction.
We compared the thermal state $\hat{\tau}_\lambda$ to the photon-subtracted thermal state $\hat{\sigma}_k=\hat{B}_k\hat{\tau}_\lambda\hat{B}^\dagger_k/\mathrm{Tr}[\hat{B}_k\hat{\tau}_\lambda\hat{B}^\dagger_k]$ and investigated the regime of $(\lambda,\eta)$ where the majorization relation $\hat{\sigma}_k\prec\hat{\tau}_\lambda$ holds.
The region is plotted from numerics in Fig. \ref{fig:num_addsub}.

The majorization relation $\hat{\sigma}_k\prec\hat{\tau}_\lambda$ ensures the existence of a LOCC to transform the photon-subracted TMSV into the original TMSV, as per Nielsen's theorem.
It notably implies that the entropy of entanglement of the photon-subtracted TMSV is greater than the one of the original TMSV, \textit{i.e.}, $S(\hat{\sigma}_k)\geq S(\hat{\tau}_\lambda)$ (where $S(\hat{\rho})=-\Tr[\hat{\rho}\ln\hat{\rho}]$ is the von Neumann entropy of $\hat{\rho}$).
For reference, we display in Fig. \ref{fig:entropy_of_entanglement} the region where the relation $S(\hat{\sigma}_k)\geq S(\hat{\tau}_\lambda)$ holds.
The quantity $S(\hat{\sigma}_k)-S(\hat{\tau}_\lambda)$ corresponds to the increase of distillable entanglement through the filtration process in the asymptotic regime.
Figure \ref{fig:entropy_of_entanglement} should then be compared to Fig. \ref{fig:num_addsub}, picturing the region where $\hat{\sigma}_k\prec\hat{\tau}_\lambda$ holds.
 As expected, whenever $\hat{\sigma}_k\prec\hat{\tau}_\lambda$ we have $S(\hat{\sigma}_k)\geq S(\hat{\tau}_\lambda)$, but the latter condition defines a wider region than the former condition as a consequence of (asymptotic)  regularization.

\section{Majorization relations among different photon-added TMSV}
\label{apd:extra_majorization_relations}

Consider photon addition of $k$ photons per mode on a TMSV state. We denote the resulting state as $|\Phi^{k,k}\rangle$. We want to compare the entanglement properties of the states $|\Phi^{k,k}\rangle$ and $|\Phi^{1,1}\rangle$ in terms of majorization relations between eigenvalues of the states $\hat{\sigma}^{k,k}$ and $\hat{\sigma}^{1,1}$, where $\hat{\sigma}^{k,k}$ is the state we get by tracing out one of the modes of $|\Phi^{k,k}\rangle$.

We get,
\begin{eqnarray}
    \hat{\sigma}^{k,k} = \sum_{n=0}^{\infty} q_n^{(kk)} |n\rangle \langle n|, 
\end{eqnarray}
where,
\begin{eqnarray}
 \label{eqapp:qnk}   q_n^{(kk)} = \frac{1}{N_{kk}} \lambda^n \binom{n+k}{k}^2
\end{eqnarray}
consist the elements of a vector denoted as $\mathbf{q}^{(kk)}$ and $N_{kk} = \sum_{n=0}^\infty \lambda^n \binom{n+k}{k}^2$ is the normalization factor.

We can write,
\begin{eqnarray}
    q_{n+1}^{(kk)}=\frac{1}{N_{kk}} \lambda^{n+1} \binom{n+k+1}{k}^2.
    \label{eqapp:qnplus1}
\end{eqnarray}
We write the following expansion,
\begin{eqnarray}
  \label{eqapp:binomial}  \binom{n+k+1}{k}^2 = \sum_{i=0}^{n+1} c_{n-i}^{(kk)} (i+1)^2
\end{eqnarray}
assuming that such coefficients $c_{n-i}^{(kk)}$ exist.
Then, using Eq. \eqref{eqapp:qnk}, Eq. \eqref{eqapp:qnplus1} can be written as,
\begin{eqnarray}
    q_{n+1}^{(kk)}&=&\frac{\lambda N_{11}}{N_{kk}}  \sum_{i=0}^{n+1} c_{n-i}^{(kk)} (i+1)^2 \frac{\lambda^i}{N_{11}} \lambda^{n-i}\\
 \label{eqapp:qnplus1v2}   &=&\frac{\lambda N_{11}}{N_{kk}}  \sum_{i=0}^{n+1} c_{n-i}^{(kk)} q_i^{(11)} \lambda^{n-i}.
\end{eqnarray}
Equation \eqref{eqapp:qnplus1v2} allows us to write,
\begin{eqnarray}
 \label{eqapp:Dvector}   \mathbf{q}^{(kk)}=\mathbf{D} \mathbf{q}^{(11)},
\end{eqnarray}
where,
\begin{eqnarray}
\nonumber    \mathbf{D} = \frac{\lambda N_{11}}{N_{kk}}
    \begin{pmatrix}
        c_{-1}^{(kk)}\lambda^{-1} & 0 & 0 & 0 & \ldots\\
        c_{0}^{(kk)} & c_{-1}^{(kk)}\lambda^{-1}  & 0 & 0 & \ldots\\
        c_{1}^{(kk)}\lambda & c_{0}^{(kk)}  & c_{-1}^{(kk)}\lambda^{-1}  & c_{0}^{(kk)}  & \ldots\\
        c_{2}^{(kk)}\lambda^{2} & c_{1}^{(kk)}\lambda  & c_{0}^{(kk)} & c_{-1}^{(kk)}\lambda^{-1}  &\ldots\\
        \vdots & \vdots & \vdots & \vdots & \ddots
    \end{pmatrix},
\end{eqnarray}
\textit{i.e.}, all columns have the same elements shifted one position down as we move from the left-hand-side to the right-hand-side ($\mathbf{D}$ is a lower-triangular circulant matrix). The non-zero elements of each column are given by the expansion coefficients of,
\begin{eqnarray}
   \left( \frac{\lambda N_{11}}{N_{kk}} \right)^{-1}= \frac{1}{\lambda}+\sum_{n=0}^\infty c_{n}^{(kk)} \lambda^n,
\end{eqnarray}
where we have assigned the value,
\begin{eqnarray}
   c_{-1}^{(kk)}=1, 
\end{eqnarray}
while the rest of the coefficients $ c_{n}^{(kk)}$ are given as the Taylor expansion coefficients of $N_{kk}/(\lambda N_{11})- 1/\lambda$ around $\lambda \to 0$, \textit{i.e.},
\begin{eqnarray}
\label{eqapp:Taylor}     c_{n}^{(kk)} = \frac{1}{n!} \lim_{\lambda \to 0}\frac{\partial^n}{\partial \lambda^n} \left[\left( \frac{\lambda N_{11}}{N_{kk}} \right)^{-1}- \frac{1}{\lambda}\right].
\end{eqnarray}
Equation \eqref{eqapp:Taylor} can ensure existence of the coefficients $c_{n}^{(kk)}$ while always ensures that each column sums to $1$. However, to prove that the matrix in Eq. \eqref{eqapp:Dvector} is column stochastic we need to prove that all its entries, \textit{i.e.}, all coefficients $c_{n}^{(kk)}$, are non-negative.

We can work out Eq. \eqref{eqapp:Taylor} some more to get,
\begin{widetext}
\begin{eqnarray}
\label{eqapp:Coeffcients}   c_{n}^{(kk)} =  \lim_{\lambda \to 0} \left[ \frac{(-1)^{n+1}}{\lambda^{n+1}}+\sum_{m=0}^n \binom{k+m}{m}^2\frac{1}{(n-m)!}     \ _2F_1(k+m+1,k+m+1;m+1;\lambda )  \frac{\partial^{n-m}}{\partial\lambda^{n-m}} \frac{(1-\lambda)^3}{\lambda (1+\lambda)}\right],
\end{eqnarray}
\end{widetext}
where $_2F_1(a,b;c;z)$ is the hypergeometric function.

For $k=2$ we compare $2$-photon addition per mode to single-photon addition per mode. For said case, utilizing Eq. \eqref{eqapp:Coeffcients}, we find,
\begin{eqnarray}
 \label{eqapp:cn22}   c_n^{(22)} = 3 n+\frac{(-1)^n}{2}+\frac{9}{2},
\end{eqnarray}
which is non-negative for all $n\geq0$, Therefore, the matrix $\mathbf{D}$ in Eq. \eqref{eqapp:Dvector} is rendered to column stochastic and consequently we obtain the majorization relation $\hat{\sigma}^{2,2} \prec \hat{\sigma}^{1,1}$. 

By setting $k=3$, we compare $3$-photon addition per mode to single-photon addition per mode. For said case, utilizing Eq. \eqref{eqapp:Coeffcients}, we find,
\begin{eqnarray}
 \label{eqapp:cn33}   c_n^{(33)} =\frac{5 n^3}{3}+10 n^2+\frac{58 n}{3}+12,
\end{eqnarray}
which is non-negative for all $n\geq0$, Therefore, the matrix $\mathbf{D}$ in Eq. \eqref{eqapp:Dvector} is rendered to column stochastic and consequently we obtain the majorization relation $\hat{\sigma}^{3,3} \prec \hat{\sigma}^{1,1}$.

For $k=4$, we compare $4$-photon addition per mode to single-photon addition per mode. For said case, utilizing Eq. \eqref{eqapp:Coeffcients}, we find,
\begin{widetext}
\begin{eqnarray}
 \label{eqapp:cn44}   c_n^{(44)} = \frac{7 n^5}{24}+\frac{175 n^4}{48}+\frac{425 n^3}{24}+\frac{125 n^2}{3}+\frac{95 n}{2}-\frac{3
   (-1)^n}{32}+\frac{675}{32},
\end{eqnarray}
\end{widetext}
which is non-negative for all $n\geq0$, Therefore, just as before, the matrix $\mathbf{D}$ in Eq. \eqref{eqapp:Dvector} is rendered to column stochastic and consequently we obtain the majorization relation $\hat{\sigma}^{4,4} \prec \hat{\sigma}^{1,1}$.

For all cases we presented (and beyond, \textit{i.e.}, up to $k=8$), we find that allowing $n=-1$, the closed form Eqs. \eqref{eqapp:cn22}, \eqref{eqapp:cn33}, \eqref{eqapp:cn44} (and similarly up to $k=8$) give $c_{-1}^{(22)}=\ldots=c_{-1}^{(88)}=1$, consistently with the value we assigned earlier. Moreover, we have verified that the coefficients we have derived using Eq. \eqref{eqapp:Coeffcients} for $k=2,\ldots,8$, are non-negative and consistent with Eq. \eqref{eqapp:binomial}. One can allow for $k \geq 9$ and explore if the matrix $\mathbf{D}$ becomes column-stochastic by proving the non-negativity of the coefficients given by Eq. \eqref{eqapp:Coeffcients}.
As $k$ grows, we observe the general trend that $c^{(kk)}_n$ becomes a polynomial of increasing degree in $n$ with mostly non-negative terms.
Although it is apparent for each computed instance of $k$ that the polynomial $c^{(kk)}_n$ is non-negative, it seems difficult to show analytically that expression \eqref{eqapp:Coeffcients} is non-negative for every $k\geq 9$.

At this point, we conjecture the non-negativity of Eq. \eqref{eqapp:Coeffcients}, \textit{i.e.}, we conjecture that the matrix $\mathbf{D}$ of Eq. \eqref{eqapp:Dvector} is column stochastic for all $k\geq 2$, implying the majorization relation $\hat{\sigma}^{k,k} \prec \hat{\sigma}^{1,1}$. In case said conjecture is invalid, it does not mean that $\hat{\sigma}^{k,k}$ and $ \hat{\sigma}^{1,1}$ do not satisfy majorization relations; it would merely mean that the methods presented in this appendix (and main paper) fail to uncover them. Therefore, our conjecture is stronger than just asserting $\hat{\sigma}^{k,k} \prec \hat{\sigma}^{1,1}$: we conjecture that $\hat{\sigma}^{k,k} \prec \hat{\sigma}^{1,1}$ \emph{and} that the column stochastic matrix is the one defined through Eqs. \eqref{eqapp:Dvector}, \eqref{eqapp:Coeffcients}.

\end{document}